\documentclass[journal, 10pt]{IEEEtran}

\usepackage{balance}
\usepackage{epsfig}
\usepackage{subfigure} 
\usepackage[nolist]{acronym}
\usepackage{nicefrac}
\usepackage{amsmath}

\usepackage{multicol}

\begin{document}

\title{TCP SIAD: Congestion Control supporting \\ High Speed and Low Latency \\ \vspace{0.2cm}
\small{ ETH TIK Technical Report 367, 23 December 2016}}

\author{
{\rm Mirja K\"uhlewind}\\
ETH Zurich, Switzerland*\thanks{*This work was performed while the author was with IKR, University of Stuttgart, Germany.}
}

\maketitle

\begin{abstract}
Congestion control has been an open research issue for more than two decades. 
More and more applications with narrow latency requirements are emerging which are not well addressed by existing proposals. 
In this paper we present TCP Scalable Increase Adaptive Decrease (SIAD), a new congestion control scheme supporting both high speed and low latency. 
More precisely, our algorithm aims to provide high utilization under various networking conditions, and therefore would allow operators to configure small buffers for low latency support.
To provide full scalability with high speed networks, we designed TCP SIAD based on a new approach that aims for a fixed feedback rate independent of the available bandwidth.
Further, our approach provides a configuration knob for the feedback rate.
This can be used by a higher layer control loop to impact the capacity share, potentially at the cost of higher congestion, e.g. for applications that need a minimum rate. 

We evaluated TCP SIAD against well-known high-speed congestion control schemes, such as Scalable TCP and High Speed TCP, as well as H-TCP that among other goals targets small buffers. 
We show that only SIAD is able to utilize the bottleneck with arbitrary buffer sizes while avoiding a standing queue. 
Moreover, we demonstrate the capacity sharing of SIAD depending on the configured feedback rate and a high robustness of TCP SIAD to non-congestion related loss.

\end{abstract}
\acresetall

\section{Introduction}

Congestion Control is an important part of the Internet since a series of congestion collapses in the late 1980s and has been an active research area since. 
Even though a large variety of proposals for congestion control exists, today most often only the default configuration of the operating system is used~\cite{Yang2011short}.
In the last 20-25 years various congestion control schemes have been proposed to either cope with certain link characteristics or specific application requirements.
Many of these proposals do not see wide deployment as application designers might not know the network conditions and system administrators do not know the application requirements.
\\
In general congestion control should
\begin{enumerate}
 \item avoid a congestion collapse,
 \item be able to utilize the bottleneck, and 
 \item support different application layer requirements. 
\end{enumerate}
While congestion avoidance is essentially provided by any control loop that reduces its sending rate based on congestion feedback,
in this paper we aim to design a congestion control scheme that addresses both of the other two by
 a) utilizing the link independent of the configured buffer size and scaling with any \ac{BDP} and 
 b) providing a configuration interface to the higher layer application that determines the aggressiveness and thereby the instantaneous share of the bottleneck capacity. 


In this paper we present TCP Scalable Increase Adaptive Decrease (SIAD), a new \ac{TCP} congestion control scheme supporting both high speed and low latency. 
Many proposals focus on high speed but only very few consider low latency, while this aspect is becoming more and more important in the Internet.
TCP SIAD provides high link utilization and fast bandwidth allocation in high speed network with arbitrary buffer sizes.
This allows network operators to configure smaller buffers and thereby reduce the queuing delay to better support low latency requirements without causing network underutilization.

One of the main contributions of TCP SIAD is a new approach called Scalable Increase that aims for a fixed feedback rate independent of the available bandwidth and thereby is fully scalable with any link speed.
A fixed feedback rate means that the feedback signal is always received with the same frequency even if the bandwidth increases.
To reach this, Scalable Increase calculates the increase rate dynamically on each congestion notification depending on the current sending rate and subsequent rate reduction.
Further, this approach introduces an additional configuration parameter, defining the number of \acp{RTT} between two congestion/loss notification events and therefore the aggressiveness.
This configuration knob can be used by a higher layer control loop to impact the capacity share at the cost of higher congestion, e.g. for applications that need a minimum rate. 

TCP SIAD's Adaptive Decrease scheme adapts the rate reduction on congestion notification to the configured network buffer size to achieve high utilization independent of the network buffer size, even if buffers are configured to be small for low latency support.
As TCP SIAD is still a loss-based scheme to be able to compete with existing loss-based congestion control in the Internet, it still needs to fuly fill the queue to receive a congestion notification. 
Therefore the maximum queuing delay is determined by the network operator buffer configuration.
However, TCP SIAD at least minimizes the average delay by avoiding a standing queue in case of large buffers. 
This provides an deployment incentive for end users even with today's buffer configuration. 

Further we introduce 
a Fast Increase phase that provides fast bandwidth allocation in dynamic high speed networks e.g when new capacity becomes available. 
While most high speed congestion control schemes address scalability by more aggressive probing at higher sending rates,
TCP SIAD increases its sending rate faster under changing network condition than in steady state.
This approach is a general concept for high speed schemes that could also be applied to other proposals.

For evaluation we implemented TCP SIAD in Linux and used this implementation within a simulation tool that integrates virtual machines into a simulated network.
The Linux implementation proves implementation feasibility. 
A controlled, simulated network environment provides best conditions for a detailed evaluation of the microscopic algorithm behavior and to show the robustness in various well-defined (extreme) scenarios that can occur in the Internet.

We show that only TCP SIAD is able to fulfill our design goals as derived and stated in the next section.
Even though TCP Cubic was not designed to target low latency support and H-TCP is not explicitly designed for high speed networks, both algorithms perform well in many scenarios.
However, only TCP SIAD is able to utilize the bottleneck with arbitrary buffer sizes while avoiding a standing queue as well as being fully scalable by inducing a fixed, configurable feedback rate that can always be reached under various network conditions.
Therefore, only TCP SIAD explicitly addresses the current research challenges summarized in the next section.
Moreover, we demonstrate 
the capacity sharing of SIAD depending on the configured feedback rate 
and a high robustness to non-congestion related losses in various extreme Internet scenarios.

\subsection{Research Challenges and Design Goals}

In this section we describe arising challenges to the current Internet and 
derive design goals from these challenges. 

\paragraph{Scalability}
Traditional congestion control such as TCP NewReno
does not scale well with high \ac{BDP} links.
TCP NewReno increases its sending rate by one packet per \ac{RTT} and halves it when congestion occurs.
Therefore this kind of scheme induces a certain loss rate which is smaller the larger the \ac{BDP} is~\cite{Padhye2000}.
First of all, for high bandwidth or high delay links, this approach is rate-limited by the theoretical limit of the network bit error rate.
Even worse, when additional capacity becomes available, this scheme needs a very long time to allocate new capacity.
E.g. raising from 5 Gbit/s to 10 Gbit/s with an \ac{RTT} of 100\,ms and 1500\,byte packets takes more than an hour.
This problem is well known and proposals to address it are discussed in section~\ref{sec:relatedWork}. 
However, most of the proposed schemes scale much better but still depend on the \ac{BDP} of the bottleneck link.
In our proposal we aim to resolve this dependency by introducing the Scalable Increase scheme that provides a fixed and configurable feedback rate independent of the link bandwidth.


\paragraph{Low Latency Support}
Further, we aim to better support emerging services that require low end-to-end latency, like real-time media or interactive cloud services. 
Often, network providers configure large buffers to enable high link utilization and low loss rates.
As loss-based congestion control always fills these buffers to induce loss as feedback signal, large buffers add high queuing delays.
That means as long as one has to compete with loss-based congestion congestion, only the buffer configuration can influence the maximum latency.
However, just changing the buffer configuration would cause link underutilization due to today's congestion control. 
Therefore only changing both, the queue management and the congestion control, would solve the problem fully.
While end-to-end congestion control cannot change the buffer configuration, it can be designed to 
a) avoid a standing queue by emptying large buffers at every decrease event and
b) keep the link utilization high even if very small buffers are configured.
Only if a solution for congestion control exists that actually can cope with small buffers as proposed in this work,
network providers can take the first step in solving the latency problem by simply configuring
small buffers (or use an AQM with a low marking threshold).
Therefore only the combination of both small buffers and the use of TCP SIAD avoid large queuing delays and still keep link utilization high.
However, even with large buffers TCP SIAD avoids a standing queue and therefore at least reduces the average latency. 

\paragraph{Per-User Congestion Policing} 
Moreover, our scheme allows the user to configure the feedback rate and thereby 
the aggressiveness.
Many proposals are complex because they are designed to be ``\ac{TCP}-friendly''. 
TCP friendliness means that a flow competing with a flow using NewReno-like congestion control achieves equal rate sharing.
However, this requirement might even prohibit the usage of 
certain services in some scenarios e.g. streaming that needs a certain minimum rate.
Therefore, we remove this requirement following the argumentation that fairness does not depend on the instantaneous rate but on the usage over time.
Fairness should be enforced on a per-user (and not per-flow) basis~\cite{Briscoe2007}.
Note, this fairness definition also allows grabbing a larger share than others (for a limited time). 
However, there is currently no mechanism that stops single flows or users to use a more aggressive congestion control and thereby push away other traffic.
In fact TCP Cubic, the default configuration in Linux, is already more aggressive than TCP NewReno.
To address this problem per-user congestion policing based on the Congestion Exposure (ConEx) protocol~\cite{rfc6789} has been proposed. 
If policing is done based on the amount of congestion that one flow is causing, the sender needs a way to control the congestion rate and thereby the aggressiveness.
Therefore to make congestion policing feasible as well as to addresses today's diverse application requirements, we aim for a congestion control design that provides an upper layer configuration interface to influence the amount of congestion caused 
and thereby the capacity sharing with other flows.

\vspace{0.2cm}

This leads to the following design goals: 
\begin{enumerate}
 \item \textbf{High link utilization:} Always utilize the bottleneck link independent of network buffer size as a precondition to configure small network buffers.
 \item \textbf{Minimize average queuing delay}: Drain the queue at every decrease to avoid a standing queue and thereby minimize the average end-to-end delay.
 \item \textbf{Fast bandwidth allocation:} Quickly allocate new available bandwidth when the network conditions have changed.
 \item \textbf{Fixed feedback rate:} Induce a fixed feedback rate when self-congested independent of the available bandwidth to scale in high speed networks.
 \item \textbf{Configurable aggressiveness}: Provide a configurable aggressiveness when competing with other traffic to support different application requirements.
\end{enumerate}



\section{Related Work}\label{sec:relatedWork}

The main task of congestion control is to determine the sending rate and thereby avoid network overload.
In most implementations congestion control is ACK-clocked: actions are taken when an acknowledgement (ACK) packet is received.
The sending rate is based on the \textit{congestion window} (cwnd) which limits the number of packets in flight.

Most existing loss-based schemes implement \ac{AIMD} as it is proven to converge.
The increase and decrease functions of the \ac{AIMD}~\cite{Jacobson1988, Chiu1989} scheme are given by 
\begin{equation}\label{eq:increase}
 cwnd = cwnd + \frac{\alpha}{cwnd} \text{ [per ACK]}
\end{equation}
\begin{equation}
 cwnd = cwnd - \beta*cwnd \text{ [on congestion event]}
\end{equation}
In loss-based \ac{TCP} congestion control, congestion is detected when 3 duplicate ACKs or an \ac{ECN} signal is received. 
For the classical TCP NewReno
congestion control scheme $\alpha$ is 1 and $\beta$ is set to 0.5 in \textit{Congestion Avoidance}.
In contrast the congestion window in \textit{Slow Start}, e.g. at the beginning of a transmission, is increased by 1 packet per ACK and therefore doubled per \ac{RTT}.
\ac{AIMD} leads to a behavior where congestion is periodically induced to probe for free capacity.
We define all losses or congestion notifications that occur in the same \ac{RTT} as one \textit{congestion event}.
The period between two congestion events is called \textit{congestion epoch} as indicated in Figure~\ref{fig:siad_1bdp_buffer}.

The average throughput of TCP NewReno can be given based on the loss probability $p$ by $\sqrt{\frac{2}{3p}}$.
Therefore the larger the bandwidth is the smaller the loss probability has to be.
The \ac{MIMD} scheme, however, increases by a fixed amount $\alpha$ (instead of $\frac{\alpha}{W}$) and therefore would solve the scalability problem but often induces high loss rates and does not guarantee convergence.

Congestion mechanisms can be categorized as delay-based and loss-based depending on feedback signal that is used to make the back-off decision.
Delay measurements give an early congestion feedback when the queue fills and induces additional delay before loss occurs.
Delay-based approaches usually 
try to limit the number of packets in the queue and thereby not only the average but also the maximum queuing delay.
Unfortunately, they can then not co-exist with loss-based mechanisms, as the delay signal triggers earlier reduction.
Therefore TCP SIAD is a hybrid scheme that used loss as the primary congestion feedback signal to decide when the sending rate is reduced but takes the delay information into account to calculate the right reaction and decide by how much the sending rate is reduced.

Further there are also proposals that are network-supported 
(such as XCP~\cite{Katabi2002short} and RCP~\cite{Dukkipati2008}) where the congested network elements decides about the capacity sharing without probing of the end-system.
We do not compare with those approaches, as they are hard to deploy. 
More recently a completely different approach called RemyCC~\cite{Winstein2013} was proposed that provides a computer-generated congestion control scheme based on a model of the network that the algorithm will be used in. 
As such a model is usually not known, especially not for the Internet, RemyCC is hard to deploy when competing with other (unknown) congestion control schemes.
Even though the approach is interesting and hopefully provides new ideas and input for congestion control design, we do not compare to this initially proposed scheme.

In the following paragraph we survey well-known 
delay-based approaches. 
Subsequently we discuss loss-based schemes for high speed networks and/or support of low latency as these are common goals with TCP SIAD.

\subsection{Delay-based Approaches}

\textbf{TCP Vegas}~\cite{Barakmo1995} 
calculates the current throughput based on the congestion window and currently measured \ac{RTT} and compares it to an expected throughput based on the base \ac{RTT} when there is no congestion.
If the difference of the actual throughput and the expected throughput is smaller than a lower threshold or larger than an upper threshold the congestion windows is linearly increased or, respectively, decreased.


\textbf{LEDBAT}~\cite{rfc6817} 
implements congestion control for background traffic that goes ``out of the way'' in the presence of loss-based, foreground traffic.
LEDBAT is based on one way delay measurements and defines a target delay 
to limit the maximum additional queuing delay that is induced by the background traffic.
Further, LEDBAT introduces a mechanism to update the base delay measurement when e.g. route changes reduce the end-to-end base delay by storing the delay minimum for each minute for the last 10 minutes. 
As TCP SIAD aims to empty the queue with every decrease, it is able to measure the base delay every congestion epoch, and thereby does inherently address the problem of changing network routes. 

\textbf{TCP FAST}~\cite{Jin2004short} is a delay-based approach that addresses the scalability problem in high speed networks by adapting the increase factor based on the difference of the current queuing delay and the target delay.
Still it does not compete with loss-based congestion control.

\subsection{High Bandwidth-Delay-Product}

\textbf{TCP Cubic}~\cite{Ha2008_short} is the default congestion control algorithm in Linux.
It uses the maximum congestion window of the previous congestion epoch 
as a reference point and calculates its sending rate dynamically between two congestion events based on a cubic function.
In addition TCP Cubic reduces the congestion window only to 0.7*cwnd (instead of halving). 
This still allows high link utilization with buffers that are smaller than the \ac{BDP} but also induces a larger standing queue otherwise, and thereby more additional queuing delay.
One of TCP Cubic's goals is to quickly resume to the previous maximum congestion window in a high speed network but then to not overshoot too much.
Consequently TCP Cubic stays for a long time at the targeted rate before starting allocating new capacity that potentially became available.
This failure in fast convergence is explicitly address by TCP SIAD's Fast Increase algorithm. 
Further, TCP SIAD also maintains a congestion window target value based on the previous maximum, but the increase and decrease functions are different.
Instead of using a fixed smaller decrease factor, TCP SIAD avoids underutilization in the first place by calculating the decrease factor dynamically.


\textbf{Scalable TCP}~\cite{Kelly2003} addresses the scalability problem by using the \ac{MIMD} scheme with $\alpha=0.01$ and $\beta=0.125$.
That means it scales better with high speed links. 
Unfortunately, it only converges if there is some randomness~\cite{Chiu1989} and often induces a high loss rate. 

\textbf{High Speed TCP}~\cite{rfc3649} is an AIMD scheme that additionally aims to be TCP-friendly at lower rates. 
It defines a congestion window threshold (of 38 packets) above which the increase gets more and more aggressive based on a target congestion feedback rate at a (large) target congestion window.


\textbf{Compound TCP}~\cite{Tan2005} adds an additional \textit{Delay Window} to the congestion window.
The Delay Window allows a higher sending rate when the bottleneck queue is empty and thus the link is underutilized.
The algorithm for the Delay Window calculation is based on TCP Vegas and estimates the amount of backlogged data in the bottleneck queue as $cwnd-\frac{RTT_{base}}{RTT}*cwnd$.
If this estimate is smaller than a threshold $\gamma$ (= 30 packets), the link is assumed to be underutilized and the Delay Window is calculated such that the scalability is comparable to High Speed TCP.
Otherwise the Delay Window is decreased until zero. 
TCP Compound still halves the congestion window on loss.

\textbf{TCP Illinois}~\cite{Liu2006} 
also uses queuing delay as an additional feedback signal. 
Other than Compound TCP, it adapts the increase $\alpha$ and decrease factor $\beta$  based on the current estimate of the queuing delay $d_a$ within one congestion epoch.
While $\beta$ increases linearly between $\beta_{min}$ and $\beta_{max}$ when the delay is between two thresholds, $\alpha$ decreases with a concave curve when the delay is above a threshold from $\alpha_{max}$ to $\alpha_{min}$ until the maximum delay is reached.
Further TCP Illinois still halves the congestion window when the queue overflows and thereby might cause underutilization in case of small queues.
In contrast TCP SIAD only adapts $\alpha$ and $\beta$ once per congestion epoch using completely different algorithm derived from different design goals.


\textbf{E-TCP}~\cite{Gu2007short} is an approach that also aims for small buffer support. 
To achieve this E-TCP defines a minimum feedback rate.
This is realized by a multiplicative increase similar to Scalable TCP, but with an increase rate of $\alpha=0.04$, and the respective decrease function to not go below a given feedback rate $p_0=0.01$.
While E-TCP resolves the scalability problem by enforcing a minimum fixed feedback rate, the approach taken by TCP SIAD is completely different and additionally allows for controlling the aggressiveness e.g. as needed for congestion policing.

\textbf{Relentless TCP}~\cite{Mathis2009} increases by one packet per ACK, similar as TCP NewReno, but decreases by one for each loss (separately).
Therefore Relentless TCP induces a fixed, very high feedback rate of one loss every two \ac{RTT} and does not allow standard TCP cross traffic to allocate any of the capacity.
Without an explicit congestion feedback signaling as \ac{ECN}, Relentless TCP is not applicable to the Internet.

\textbf{H-TCP}~\cite{Leith2004} is 
based on queue length estimation 
as introduced by TCP Vegas and also used by Compound TCP but uses the estimated queue length to adapt the decrease factor $\beta$. 
However, H-TCP restricts $\beta$ to [0.5, 0.8].
This allows H-TCP to utilize links with smaller queues (to certain degree) but still causes a standing queue when the buffer is large.
Further, this approach, and thereby H-TCP, only completely drains the queue if all flows are synchronized. 
For the increase, H-TCP maintains 
a low speed mode where $\alpha$ is calculated based on the current $\beta$ for TCP-friendliness and an high speed mode where it is aligned to High Speed TCP's calculation.
TCP SIAD implements the same $\beta$ calculation than H-TCP.
However, to address the case where flows are not synchronized, which is the usual case for larger aggregates, TCP SIAD also performs additional decreases in case the queue could not be drained completely at first.
Further, TCP SIAD does not restrict $\beta$, as the proposed increase calculation partly compensates by dynamically calculating a higher increase rate in case of a wrong delay estimate and, subsequently, a too strong rate reduction.

In summary, TCP SIAD maintains a target value which is a similarity to TCP Cubic but implements a new and completely different increase function, called Scalable Increase, that aims for a fixed feedback rate.
TCP SIAD's decrease function is the same than used by H-TCP but without the limitations of H-TCP and further extended its by a new mechanism, called Additional Decrease, to handle non-synchronized reduction and still drain the queue. 
In our evaluation we compare TCP SIAD to both of these two approaches as well as Scalable and High Speed TCP as these are the most well-known implemented high speed approaches.

\subsection{Low Latency}
\ac{DCTCP}~\cite{Alizadeh2010a} is designed for low latency support in data centers.
It can only be used in a controlled environment 
as it is based on \ac{ECN} and a specific \ac{AQM} configuration. 
Recent work also considered the deployment of \ac{DCTCP} in the Internet with co-existence of other TCP traffic, but still implies changes in the network elements~\cite{Kuehlewind2014}.

In the IETF currently a working group focuses on RTP Media Congestion Avoidance Techniques (rmcat) where one requirement is low delay.
Current proposals~\cite{draft-johansson-rmcat-scream-cc, Zhu2013} are delay-based and not yet able to compete with loss-based cross traffic.
As we take a loss-based feedback signal as precondition to compete with existing traffic, we do not compare to these proposals.

\section{Algorithm Design}\label{sec:algo}

The proposed algorithm design is named TCP SIAD as composed of two basic components
\begin{itemize}
 \item \textbf{Scalable Increase} and
 \item \textbf{Adaptive Decrease}.
\end{itemize}

In contrast to most {AIMD}-based proposals, for TCP SIAD the increase step $\alpha$ and the decrease factor $\beta$ are not fixed but dynamically re-calculated once in each congestion epoch. 
However, the increase function within the congestion epoch is still linear with a increase of $\alpha$ packets per \ac{RTT}, and a multiplicative decrease of $\beta*cwnd$ is performed. 
Therefore TCP SIAD is still an \ac{AIMD} scheme.

Scalable Increase calculates $\alpha$ such that the distance between two congestion events is the same in each epoch (fixed feedback rate). 
Adaptive Decrease reduces the congestion window on congestion notification by the estimated number of back-logged packets in the queue to avoid underutilization as well as unnecessary additional queuing delays. 


\begin{figure}[t]
\centering
\subfigure[E.g. 1 BDP buffer.]{
  \includegraphics[width=0.22\textwidth]{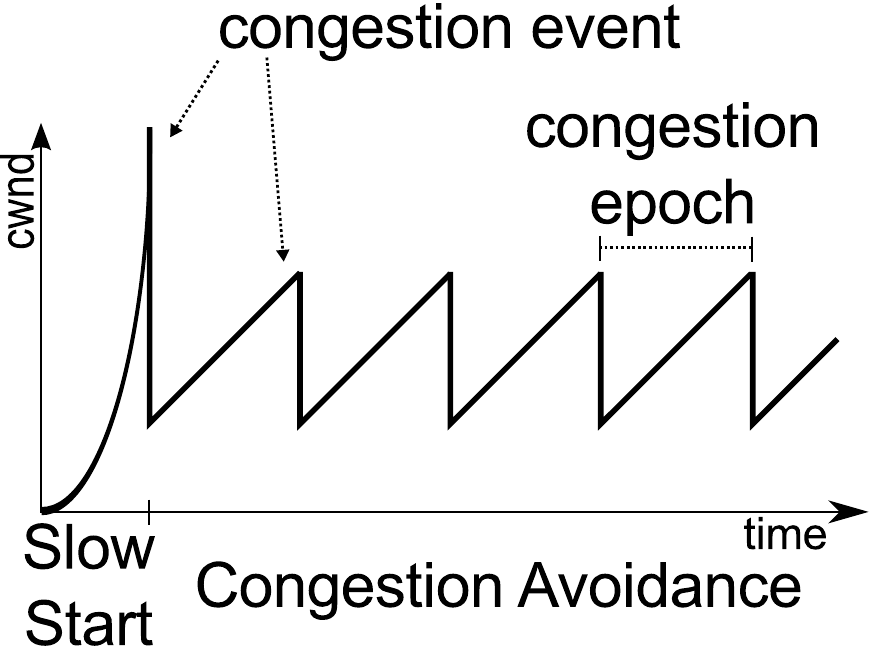}
  \label{fig:siad_1bdp_buffer}
}
\subfigure[Smaller buffer.]{
  \includegraphics[width=0.22\textwidth]{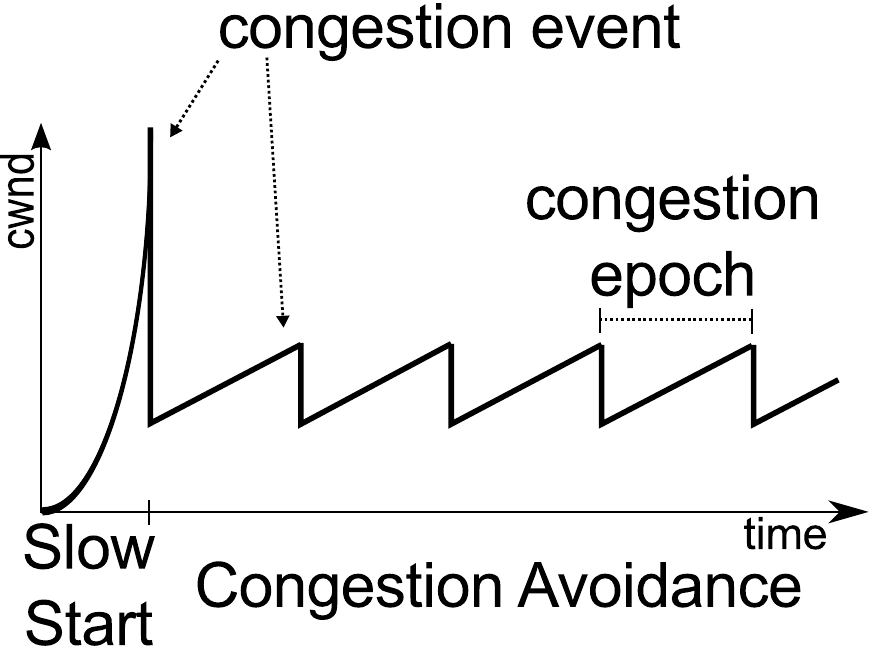}
  \label{fig:algo:siad_small_buffer}
}
\caption{The basic SIAD principle: Adapt decrease factor to queue size and increase rate to maintain the configured congestion epoch.}
\end{figure}

Figure~\ref{fig:siad_1bdp_buffer} shows that TCP SIAD with one \ac{BDP} of network buffering and an appropriately configured feedback rate looks similar to TCP NewReno.
In Figure~\ref{fig:algo:siad_small_buffer} for a reduced buffer size, Adaptive Decrease calculates a smaller decrease factor to keep link utilization high and Scalable Increase calculates a smaller increase rate to maintain the same congestion feedback frequency than in the scenario with larger buffer.

\begin{figure}[t]
\centering
\includegraphics{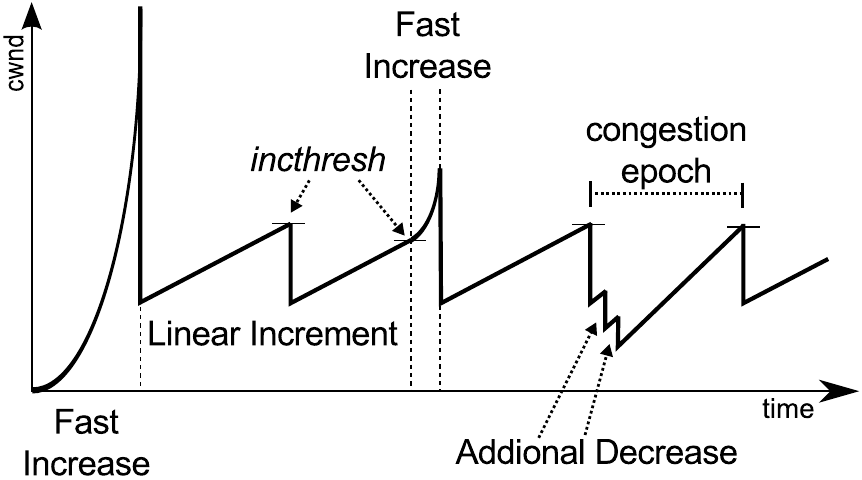}
\caption{Congestion window development of TCP SIAD.}
\label{fig:algo:siad}
\end{figure}

This basic principle of SIAD is complemented by three extensions which are sketched in Fig.~\ref{fig:algo:siad}:
\begin{description}
 \item[\textbf{Additional Decrease}] \hfill \\
 One or more Additional Decreases can be performed during a congestion epoch, not only on congestion notification.
 Additional Decrease aims to empty the queue completely at least once in a congestion epoch in case the regular decrease was not able to, due to e.g. competing traffic or measurement errors.
 \item[\textbf{Fast Increase}] \hfill \\
 In Congestion Avoidance we introduce two phases, \textit{Linear Increment} and \textit{Fast Increase}.
 When the congestion window grows above the Linear Increment threshold $incthresh$, which is also the target value for the $\alpha$ calculation of Scalable Increase, TCP SIAD enters the \textit{Fast Increase} phase.
 In this phase there is no target value for the congestion window known as new capacity is likely to be available.
 Therefore TCP SIAD increases the sending rate exponentially to quickly allocate new bandwidth.
 Slow Start is also implemented based the same increase function as used in Fast Increase using appropriate initial values for $incthresh$ and $\alpha$.
 \item[\textbf{Trend}] \hfill \\
 We calculate the target value $incthresh$ such that it depends on the maximum value of the congestion window at the current congestion event as well as the previous maximum value. 
 By this approach we introduce a trend that influences the increase factor $\alpha$ in dynamic scenarios and thereby improves convergence.
\end{description}

\subsection{Scalable Increase Adaptive Decrease (SIAD)}

TCP SIAD is composed out of five algorithms: Scalable Increase, Adaptive Decrease, Trend calculation, Fast Increase, and Additional Decrease.
Scalable Increase and Adaptive Decrease re-calculate $\alpha$ and $\beta$ for each congestion notification, and therefore once per congestion epoch.
In addition $\alpha$ is re-calculated after each Additional Decrease and continuously adapted in Fast Increase.
The detailed algorithms are described in this section while discussions on the trend calculation, convergence, RTT estimation, as well as implementation complexity are provided in subsequent sections.

\subsection*{Scalable Increase}

Scalable Increase calculates the increase step $\alpha$ as
\begin{equation}\label{eq:algo:alpha}
 \alpha = \frac{incthresh - ssthresh}{Num_{RTT}}, 1 < \alpha < ssthresh
\end{equation}
with $ssthresh$ being the Slow Start threshold, which is congestion window after the window reduction, and $incthresh$ the target value to 
be reached at the end of the next congestion epoch.
$Num_{RTT}$ gives the number of \acp{RTT} between two congestion events. 
This parameter is configurable and indicates TCP SIAD's desired aggressiveness.
Alternatively, $Num_{ms}$ can be set to the length of a congestion epoch in milliseconds.
In this case $Num_{RTT}$ is consequently estimated based on $Num_{ms}$ and the expected average \ac{RTT} for the next congestion epoch.
$\alpha$ is limited to a minimum increase of 1 packet per \ac{RTT} similar to TCP Reno in Congestion Avoidance and a maximum exponentially increase rate as in Slow Start.

Within one congestion epoch we linearly increase by the standard Additive Increase function as given in Eq.~\ref{eq:increase}.
Therefore the given calculation in Eq.~\ref{eq:algo:alpha} implements a constant feedback frequency as the total number of packets that the congestion window should be increased within one congestion epoch ($incthresh - ssthresh$) is divided by the desired number of \acp{RTT} in one congestion epoch.



\subsection*{Linear Increment Threshold and Trend}
The Linear Increment threshold $incthresh$ is calculated 
by
\begin{equation}
incthresh = cwnd_{max} + trend, incthresh \ge ssthresh
\end{equation}
where $cwnd_{max}$ is the estimated congestion window when the congestion occurred as described further below in Sec.~\ref{sec:algo:AD}.
Further $trend$ is calculated by
\begin{equation}
trend = cwnd_{max} - prev\_cwnd_{max} 
\end{equation}
where $prev\_cwnd_{max}$ is the estimated maximum congestion window at the previous congestion event.
Therefore $trend$ can be positive or negative.
However, the new target value $incthresh$ cannot be smaller than the congestion window after the decrease $ssthresh$.

%

\subsection*{Fast Increase}


If the congestion window reaches its target value $incthresh$, TCP SIAD enters the \textit{Fast Increase} phase assuming that new capacity has become available.
However, as we do not have a new target value, we carefully probe around the old target value and increase the sending rate more and more aggressively over time.
More precisely, as soon as the congestion window value reaches $incthresh$, $\alpha$ is first reset to 1 and subsequently above $incthresh$ (and below the Slow Start threshold $ssthresh$) $\alpha$ is doubled once per \ac{RTT} by calculating
\begin{equation}
 \alpha \leftarrow \alpha + \frac{\alpha}{cwnd} \text{ [per ACK]}, \alpha \le \frac{cwnd}{2}.
\end{equation} 
In Fast Increase we limit $\alpha$ to a maximum value of $\frac{cwnd}{2}$ and thereby increase the congestion window by not more than 1.5$*cwnd$ per \ac{RTT}.
This avoids too large oscillation and achieves a more stable behavior.


In Slow Start (and Linear Increment), however, the maximum value is $\alpha=cwnd$.
Setting $incthresh$, $\alpha$ as well as $cwnd_{max}$ to the initial congestion window value at the start of a connection leads to an exponential increase that doubles its rate every \ac{RTT} as desired for Slow Start.
When Slow Start is left and there is no valid $incthresh$ yet (that is larger than $ssthresh$), we reset $\alpha$ to 1 and enter Fast Increase directly.
If there is an $incthresh$ that is larger than $ssthresh$, we calculate $\alpha$ as in Eq.~\ref{eq:algo:alpha} and enter Linear Increment.

\subsection*{Adaptive Decrease}\label{sec:algo:AD}

Adaptive Decrease reduces the congestion window on congestion notification by the number of back-logged packets in the queue $q$.
As used by TCP Vegas and H-TCP, $q$ is estimated as
\begin{equation}\label{eq:algo:qlenght}
 q = \frac{RTT_{curr}-RTT_{min}}{RTT_{curr}} * cwnd.
\end{equation}
Note, for $RTT_{curr}$ we use the minimum of the last two \ac{RTT} measurement samples before the congestion event to filter out single outliers.
Consequently Adaptive Decrease sets the congestion window to
\begin{equation}
 cwnd = cwnd_{max} - q - 1 = \beta * cwnd_{max} - 1  \text{ [on congestion]}
\end{equation}
with a multiplicative decrease factor $\beta$ of
\begin{equation}
 \beta = \frac{RTT_{min}}{RTT_{curr}}.
\end{equation}
We additionally decrease by one, as it is important for us to ensure that the queue drains completely.
However, the congestion window is cropped to a minimum value of $MIN\_CWND=2$.

$cwnd_{max}$ is calculated based on the current congestion window value $cwnd$ (before the decrease) minus the number of increases that where performed in the last \ac{RTT}. 
We perform this additional adaptation as the congestion feedback has a delay of one \ac{RTT}.
During this time period the congestion window increases further depending on which operation phase we are currently in.
However, to apply the correct window reduction such that the queue drains completely, we must estimate the congestion window value at the time when the congestion has happened (about one \ac{RTT} ago).

In Linear Increment the congestion window is increased by $\alpha$ packets per \ac{RTT}.
In Fast Increase not only $cwnd$ is increased but also the increase step $\alpha$ itself changes.
And in addition we have to consider the cases where we switch from 
one phase to the other just during the last \ac{RTT} separately.
This leads to five different cases:
\begin{enumerate}
 \item \textbf{Default in Linear Increment:} For the default case in Linear Increment we reduce the current window value by $\alpha$ (the number of increases per \ac{RTT}): \\
   \begin{equation}\label{eq:algo:maxcwnd1}
   cwnd_{max} =  cwnd - \alpha 
   \end{equation}
 \item \textbf{In Slow Start or Fast Increase: } In Slow Start or Fast Increase (far enough above the Linear Increment threshold) we reduce the current window by $\alpha/2$ as $\alpha$ was doubled during the last \ac{RTT}: \\
   \begin{equation}\label{eq:algo:maxcwnd2}
   cwnd_{max} =  cwnd - \frac{\alpha}{2}
   \end{equation}
 \item \textbf{Linear Increment threshold passed: }  If the Linear Increment threshold was just passed during the last \ac{RTT}, $\alpha$ has been reset to 1. However, we still have to apply the increase step that was used before in Linear Increment. Therefore we need to calculate this value based on the Linear Increment threshold $incthresh$ and the Slow Start threshold $ssthresh$ (similar as in Eg.~\ref{eq:algo:alpha}): \\
   \begin{equation}\label{eq:algo:maxcwnd3}
   cwnd_{max} =  cwnd - \frac{incthresh-ssthresh}{Num_{RTT}}
   \end{equation}
 \item \textbf{Maximum increase rate in Fast Increase:}  If in Fast Increase the maximum increase rate of $\alpha = \frac{cwnd}{2}$ was already reached, we reduce the congestion window by one third: \\
   \begin{equation}\label{eq:algo:maxcwnd3}
   cwnd_{max} =  cwnd - \frac{cwnd}{3}
   \end{equation}
 \item \textbf{Maximum increase rate in Slow Start:} In Slow Start, however, the maximum value is $\alpha = cwnd$. Therefore we halve, respectively: \\
   \begin{equation}\label{eq:algo:maxcwnd4}
   cwnd_{max} =  cwnd - \frac{cwnd}{2}
   \end{equation}
\end{enumerate}


For simplification, we neglect the rare case when the Slow Start threshold was just passed as one should not be too conservative in this case.
Note that we do not apply Eg.~\ref{eq:algo:alpha} if a congestion notification occurs right after a window reduction.
If the congestion window has not been increased yet, there is no adjustment is needed.

\subsection*{Additional Decrease}

Additional Decreases are performed immediately within a congestion epoch without further (explicit) congestion notification.
Additional Decrease is applied if the regular decrease (Adaptive Decrease) was not able to drain the queue fully.
We assume that the queue is not empty if we cannot measure the minimum \ac{RTT} within the next \ac{RTT} after the window reduction.
This might happen if the competing flows are not synchronized and/or use a different decrease scheme. 

However, in this situation a single flow cannot know what the right decrease factor must be.
Therefore, we allow for multiple Additional Decreases during one congestion epoch.
As a congestion epoch is $Num_{RTT}$ \acp{RTT} long and at most one Additional Decrease is performed per \ac{RTT}, we can at maximum perform $Num_{RTT}-1$ Additional Decreases per congestion epoch.
This leaves at least one remaining \ac{RTT} to finally increase to the target value.
Of course after each Additional Decrease, $\alpha$ has to be recalculated to still achieve the desired target value within the remaining time of the congestion epoch.

Each Additional Decrease is calculated the following way: \\
1) The congestion window is adapted similar as in Adaptive Decrease by
\begin{equation}
 cwnd = \frac{RTT_{min}}{RTT_{curr}} ssthresh - 1
\end{equation}
where in this case $ssthresh$ is the congestion window value about one \ac{RTT} ago as an Additional Decrease is performed about one \ac{RTT} after the previous decrease. 
This calculation decreases the congestion window only slightly.
In Additional Decrease the current \ac{RTT} $RTT_{curr}$ is still larger than the minimum as the queue is not empty, however, this does not decrease enough to actually drain the queue.\\
2) The congestion window is further reduced to
\begin{equation}
 cwnd \leftarrow max(cwnd - max(red, \alpha_{new}), MIN\_CWND)
\end{equation}
where $\alpha_{new}$ is the new $\alpha$ that we might apply after the reduction and $red$ is a reduction factor that is intended to, in the worst case, reduce the congestion window to its minimum ($MIN\_CWND=2$) over the remaining \acp{RTT} of the current congestion epoch.

$red$ is calculated by
\begin{equation}
 red = cwnd * \frac{1}{Num_{RTT}-cnt_{dec}}
\end{equation}
where $cnt_{dec}$ is the number of additional decreases that have already been performed during this congestion epoch (including the current one). 
If the competing traffic is decreasing less than TCP SIAD in Adaptive Decrease, there remains a standing queue (even after a synchronized reduction).
In this case the maximum that TCP SIAD can do to reduce the queuing delay is to set $cwnd$ to its minimum.
Even in this (extreme) case TCP SIAD is still able to grab its desired share of the network capacity, as Scalable Increase calculates a higher sending rate in response.
Note that we do not perform any further Additional Decreases if $cwnd$ is already set to $MIN\_CWND$.

In addition we have to ensure that the congestion window is at least decreased by 
\begin{equation}
\alpha_{new}=\frac{incthresh - cwnd}{Num_{RTT}-cnt_{dec}-1}.
\end{equation}
$\alpha_{new}$ is the new number of packets by which the congestion window will be increase in the next \ac{RTT}. 
However, during this \ac{RTT} we would like to keep the queue empty to be able to measure the minimum \ac{RTT}. 


If $red$ is larger than $\alpha_{new}$, Scalable Increase re-calculates $\alpha$ by
\begin{equation}
\alpha=\frac{incthresh - cwnd}{Num_{RTT}-cnt_{dec}}.
\end{equation}
Otherwise we set $\alpha=\alpha_{new}$.
Further if the new $\alpha$ is larger than the current congestion window, we set to its maximum value $\alpha=cwnd$ and do not perform any further Additional Decrease anymore. 

Finally, we set $ssthresh$ to $cwnd-1$.

\subsection{Design Choices and Reasoning}

This section provides some reasoning and theoretical background 
but does not represent a comprehensive discuss of the full design space due to space limitations.

The average throughput of Scalable Increase and Adaptive Decrease (only) in steady state can be derived using the same approach as given in~\cite{Mathis1997} to be
\begin{equation}
B(p)=\frac{2}{(\beta+1)p}
\end{equation}
where $p$ is the probability that a congestion event appears.
TCP SIAD therefore provides full \textbf{scalability} as its throughput depends on $\frac{1}{p}$ and not $\frac{1}{\sqrt{p}}$. 

\ac{AIMD} has been shown to converge to equal capacity sharing for competing flows with the same aggressiveness from any starting point~\cite{Chiu1989}. 
This is because the multiplicative decrease leads to a smaller decrease for the flow(s) with the smaller sending rate(s).
Therefore if all flows use the same increase rate, the smaller rate flows end up at a higher congestion window at the next congestion event which then converges to equal sharing. 
With Scalable Increase, unfortunately, the flow with the smaller absolute decrease also calculates a smaller increase rate as it targets the same maximum congestion window value than before.
Therefore the basic principle of Scalable Increase Adaptive Decrease provides very stable operation at a certain sharing ratio but no convergence to equal sharing.
To re-introduce equal sharing (when operating under the same aggressiveness) we add a trend value to the target value. 
Any deviation from the above described principle re-introduce \textbf{convergence} (even random values).
We decided to calculate trend in a way as described in the previous section to additionally speed up convergence by amplifying the trend that was observed between the current and the previous maximum congestion window values. 
We also experimented with other approaches, however the improvement did not justify further complexity.
Moreover there is always a trade-off between responsiveness for (fast) convergence and smoothness regarding window oscillations that is addressed well by the current approach.

Even with the introduction of trend that leads to slightly higher oscillation, TCP SIAD is more robust to \textbf{\ac{RTT} variation} and permanent changes (e.g. due to route changes) than most other hybrid schemes.
Usually these schemes, including H-TCP, do not update their $RTT_{min}$ estimate 
but use the absolute minimum \ac{RTT} value that could be observed during the connection.
TCP SIAD aims to measure the minimum after every decrease.
Therefore TCP SIAD updates in each epoch, either when an \ac{RTT} sample that is smaller than the current $RTT_{min}$ or the same sample value can be observed in subsequent \acp{RTT}.
In the latter case the queue must be empty, as the \ac{RTT} did not increase even though TCP SIAD increased its sending rate during the measurement period.
If we cannot observe nor update the current $RTT_{min}$, Additional Decrease performs further decreases under the expectation that the minimum can be observed afterwards.

In general 
we decided to rather decrease too much than causing unnecessary delays.
In contrast, H-TCP~\cite{Leith2004} aims to avoid too strong underutilization by only allowing a decrease of 0.3-0.5\,$*cwnd$ but might cause a standing queue especially as competing flows are most often not synchronized. 
In return TCP SIAD might decrease too strongly due to Additional Decrease and consequently underutilize the link.
However, Scalable Increase is designed such that it calculates an even larger increase factor to still reach the target value in time. 
Therefore \textbf{underutilization} usually only occurs for a short time, 
in contrast to schemes with a fixed increase rate.

\subsection{Implementation and Complexity}

We implemented TCP SIAD in Linux 3.5.7.
This kernel implements an initial congestion window of 10 packets as well as Proportional Rate Reduction (PPR)~\cite{rfc6937}.
PRR guarantees that the congestion window is set to the newly calculated Slow Start threshold $ssthresh$ at the end of the Fast Recovery phase. 
Note that delayed ACKs in Linux lead to an increase rate of only 1/2 packet per RTT for NewReno (as Appropriate Byte Counting (ABC) is not supported anymore).
For TCP SIAD, we estimated the number of acknowledged packets based on the newly acknowledged bytes and the Maximum Segment Size (MSS) and take this into account for the increase.

Depending on the use of the TCP Timestamp option (TSOpt) Linux provides either one \ac{RTT} sample per ACK or a smoothed average of \ac{RTT} samples which leads to a certain inaccuracy but still allows TCP SIAD to work.

A default value for $Num_{RTT}$ can be set at module loading time as well as 
later on by using a new sysctl \textit{net.ipv4.tcp\_siad\_num\_rtt} or even during the transmission using a new \ac{TCP} socket option called \textit{TCP\_SIAD\_NUM\_RTT}.

To avoid timer handling, we measure the (minimum) \ac{RTT} when the congestion window reaches $ssthresh+\alpha+1$ 
and finally make the Additional Decrease decision after the next increase. 

As most of the processing is only done on congestion notification, the computation effort is not higher than for most other schemes. 
Note, if further computational optimization is needed, the number of divisions could be strongly reduced by only allowing powers of 2 for $Num_{RTT}$. 

\begin{figure*}[t]
\centering
  \subfigure[TCP NewReno.]{
    \label{fig:reno}
    \includegraphics[trim = 10mm 20mm 10mm 15mm, clip, width=0.4\textwidth]{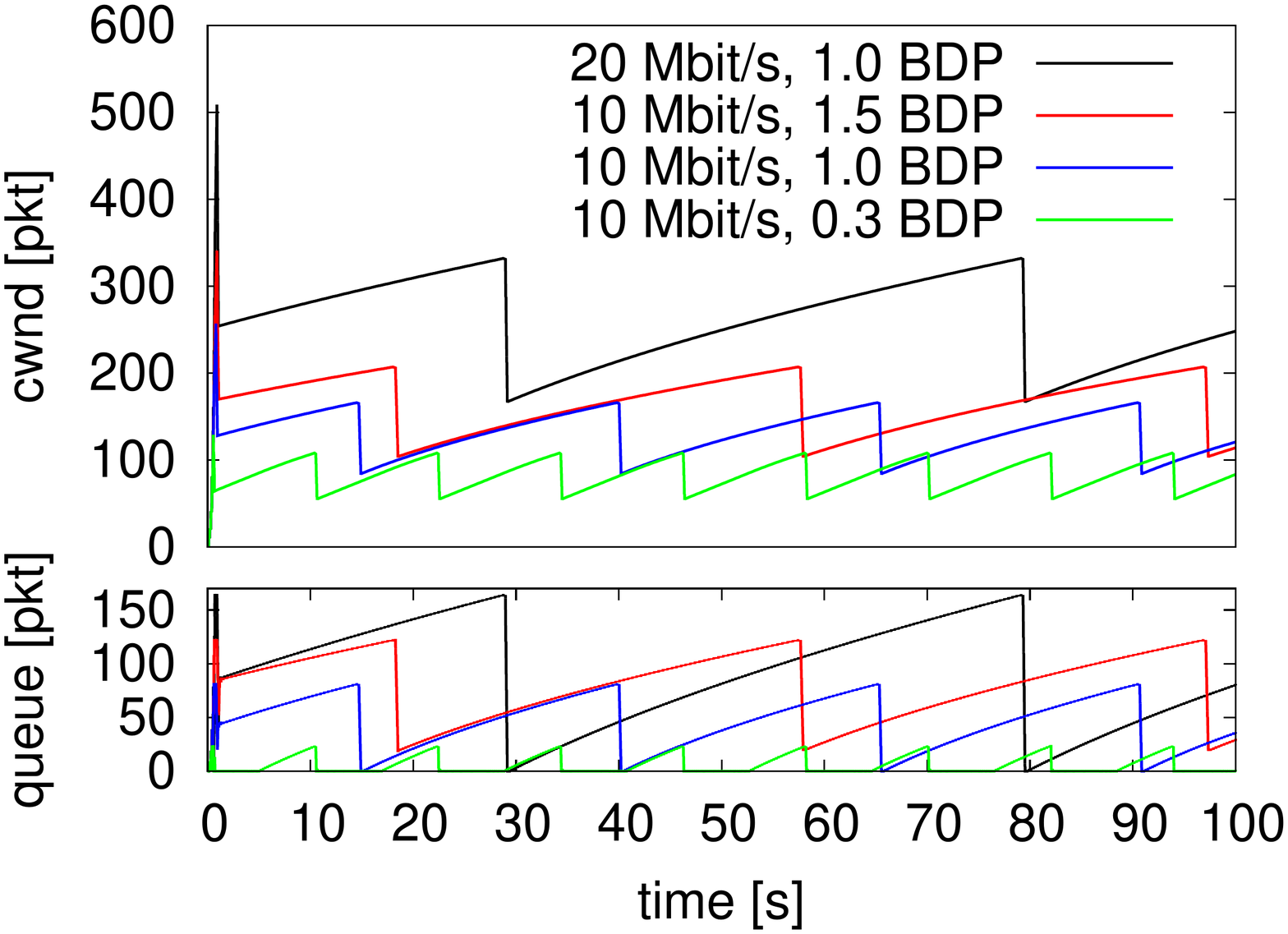}
    }
  \subfigure[TCP Cubic.]{
    \label{fig:cubic}
    \includegraphics[trim = 10mm 20mm 10mm 15mm, clip, width=0.4\textwidth]{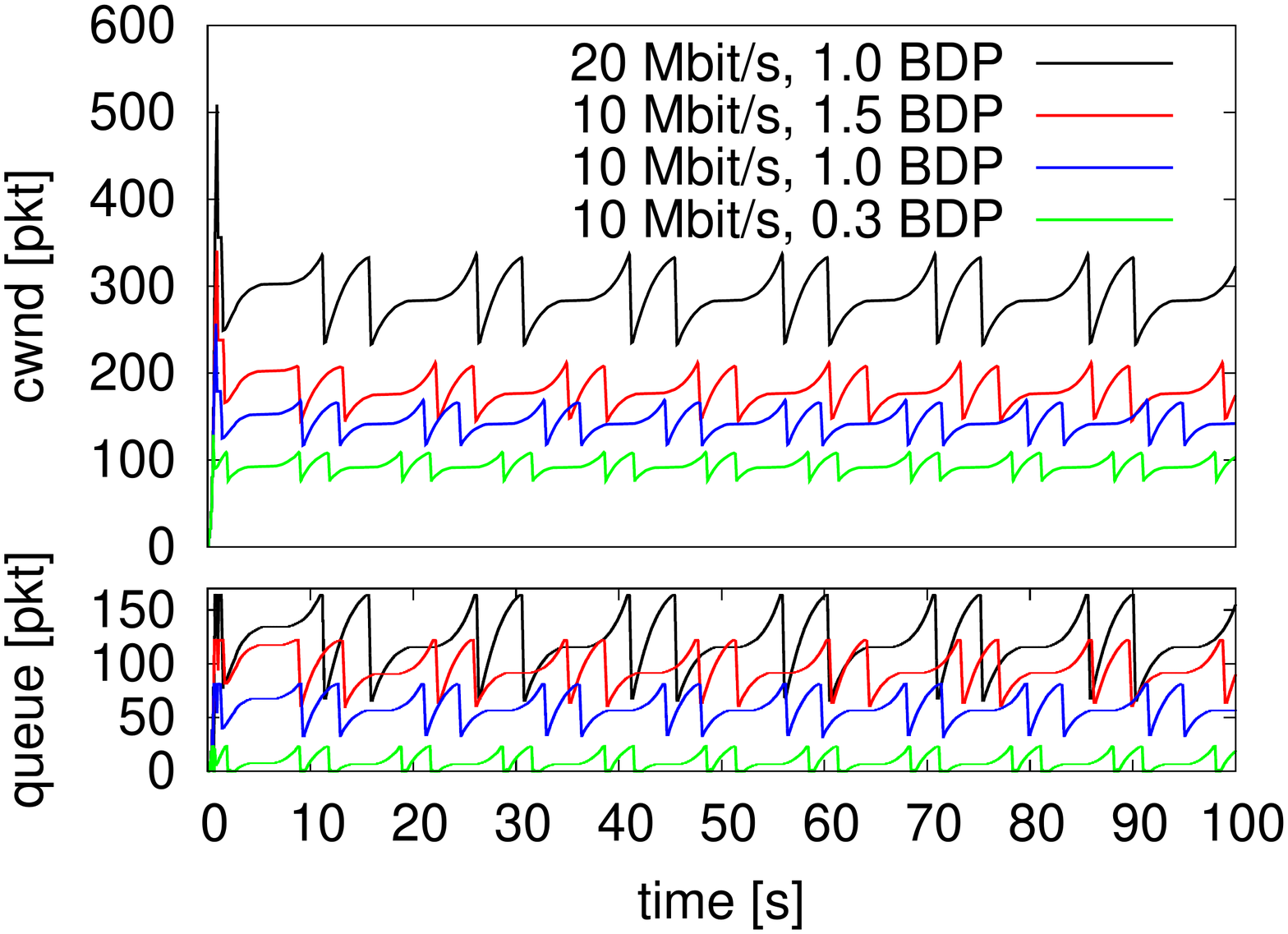}
    }
\\

   \subfigure[HighSpeed TCP.]{
     \label{fig:highspeed}
     \includegraphics[trim = 10mm 17mm 10mm 15mm, clip, width=0.4\textwidth]{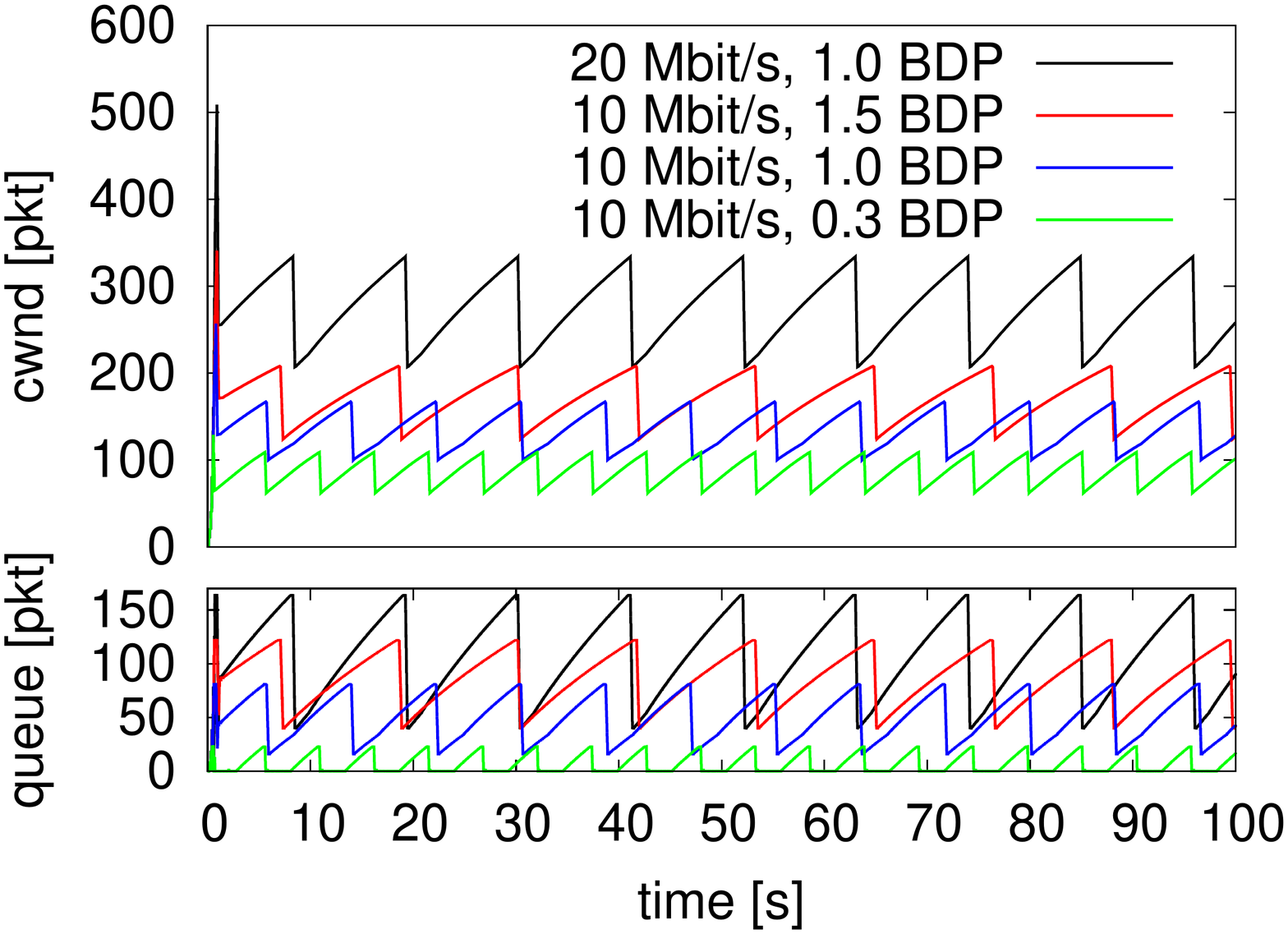}
     }
   \subfigure[Scalable TCP.]{
     \label{fig:scalable}
     \includegraphics[trim = 10mm 17mm 10mm 15mm, clip, width=0.4\textwidth]{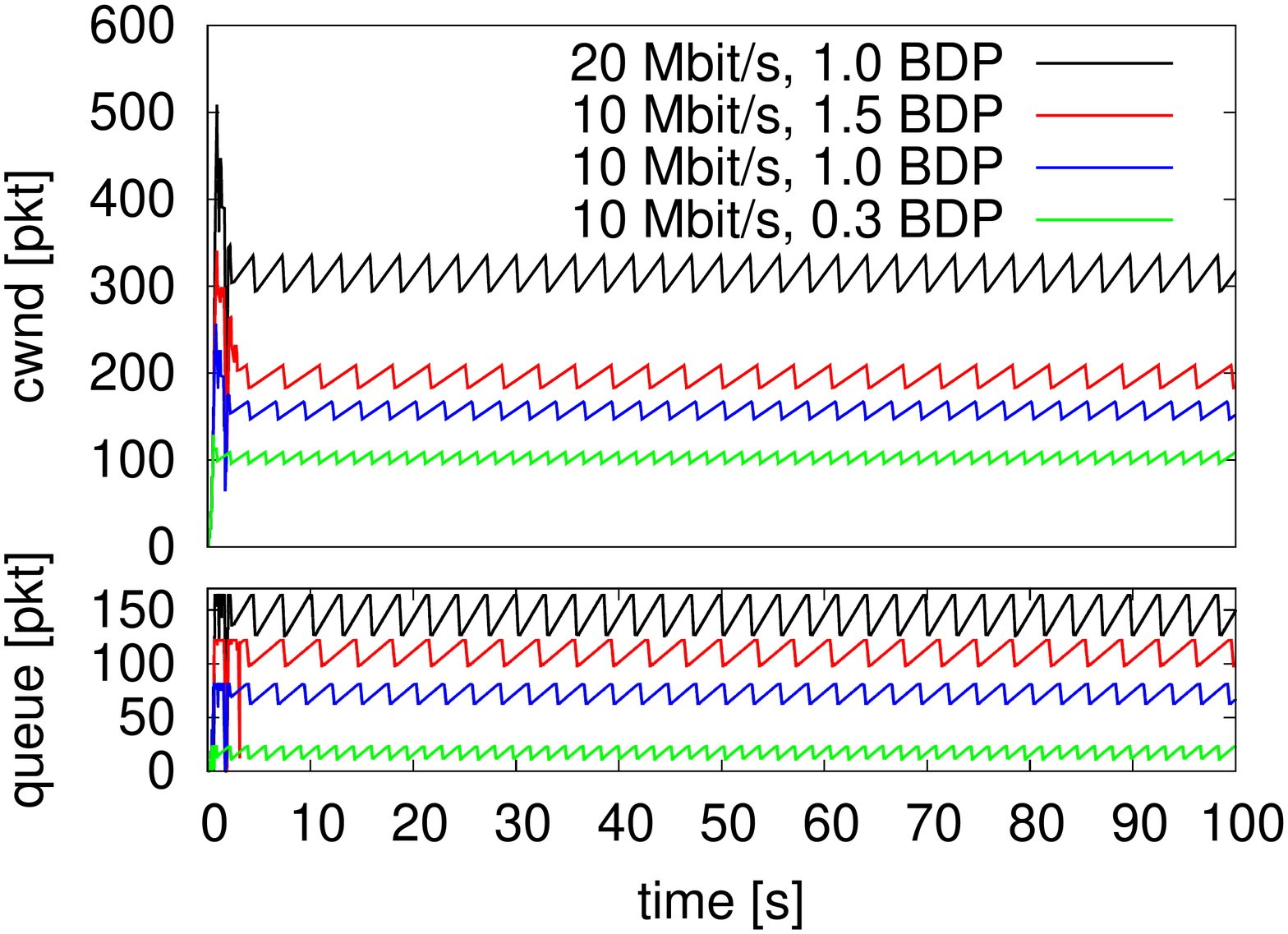}
     }
\\


  \subfigure[H-TCP.]{
    \label{fig:htcp}
    \includegraphics[trim = 10mm 17mm 10mm 15mm, clip, width=0.4\textwidth]{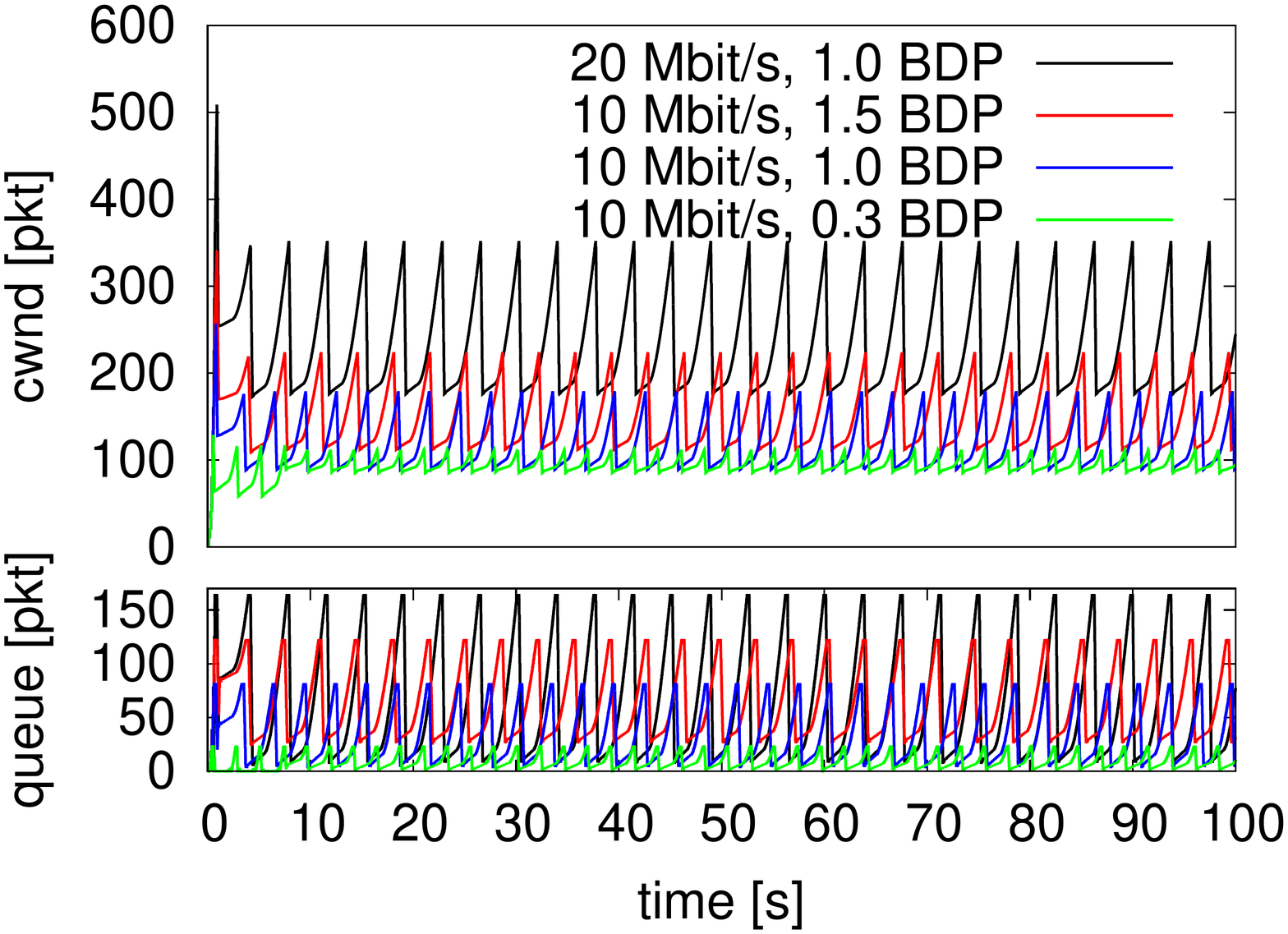}
    }
  \subfigure[TCP SIAD ($Num_{RTT}=20$).]{
    \label{fig:siad}
    \includegraphics[trim = 10mm 17mm 10mm 15mm, clip, width=0.4\textwidth]{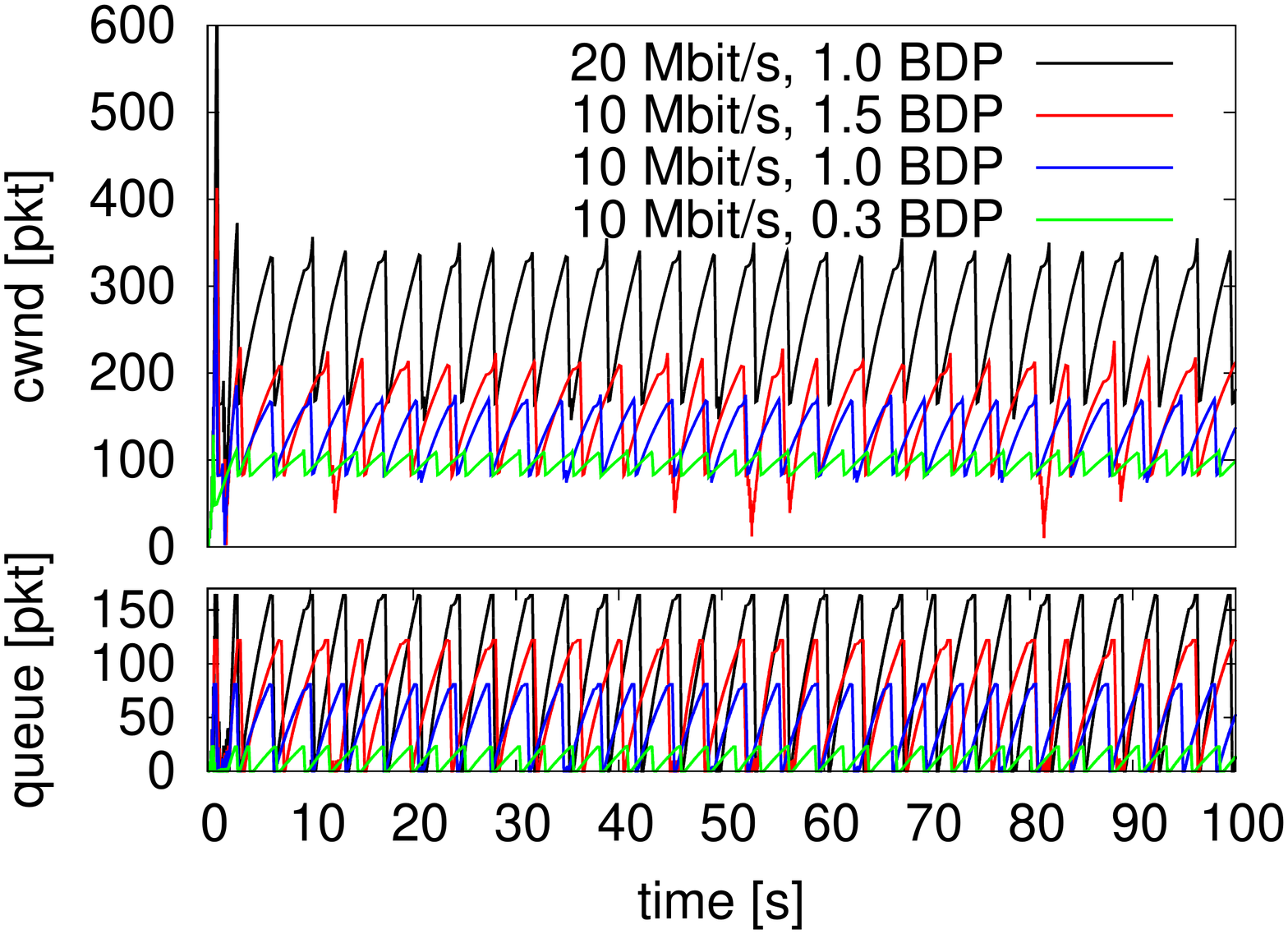}
    }
\caption{Congestion window and queue length over time for a single flow using different congestion control schemes.}
\label{fig:cwnd1flow}
\end{figure*}

\section{Evaluation}\label{sec:eval}

In this section we present an evaluation of TCP SIAD in comparison to other proposals as well as in selected extreme scenarios. 
Our evaluation is based on the event-driven network simulator IKR SimLib~\cite{SimLib} with integration for virtual machines to use unmodified Linux kernel code (VMSimInt)~\cite{Werthmann2014} which thereby provides realistic TCP behavior.
We use our Linux implementation as well as the existing Linux implementations of the other schemes. 
By using a simulated network instead of a real testbed, we are able to fully control the network conditions and therefore eliminate unwanted sources of irritation.
We use the default Linux configuration with Selective Acknowledgment (SACK) and TSOpt enabled.
We do not use the ns-3 simulation environment as IKR VMSimInt provides comparable results~\cite{Werthmann2014} and allows an easy use of unmodified and recent kernel code. 


In our simulation we used four virtual machines connected over a simulated network. 
Two machines act as senders connected to the bottleneck link and two as receiver with a symmetric back channel for acknowledgements.
This set-up allows the use of different congestion control schemes on each sender machine where each sender can have one or more connections.
At the bottleneck link, the bandwidth and the size of a tail drop buffer are configurable.
In most cases it is sufficient to vary the bandwidth or delay to evaluate different \acp{BDP}. 
Therefore we choose the same (symmetric) \ac{OWD} of 50\,ms for all presented results as an average value for the Internet. 
We vary the bandwidth from 1 to 100\,Mbit/s to cover a wide range of access link capacities.
The queue size is always represented as a multiple of the base \ac{BDP} (when the queue is empty).

At the receiver side we measure the per-flow rate, loss probability as well as loss event distance. 
The loss event distance is the time between the last loss of the previous congestion event and the first loss of the current congestion event.
All simulations ran for 600 seconds.
For the following statistical evaluations the first 20 seconds were not considered 
as we only evaluate steady state behavior. 
Therefore it is uncritical if the start-up phase is too long but we have to ensure that it is long enough to exclude Slow Start and the initial convergence phase.
Convergence is evaluated separately in section~\ref{sec:eval:multipleFlows}.
%

\subsection{Basic Analysis with a Single Flow}\label{sec:Single}

In this scenario we evaluate only one greedy connection between of one the sender/receiver pairs.
Figure~\ref{fig:cwnd1flow} shows the congestion window  and queue length over time for several simulation runs with different bandwidth and queue size configurations to characterize the different (high speed) congestion control schemes.
As a reference Figure~\ref{fig:reno} shows the behavior of TCP NewReno. 
It can be seen that TCP NewReno's feedback rate strongly depends on the \ac{BDP} and the maximum queue length.
Moreover, the queue runs empty if the configured buffer size is smaller than the base \ac{BDP} and therefore the link cannot be fully utilized by TCP NewReno.
In contrast, as shown in Figure~\ref{fig:cubic}, TCP Cubic can utilize bottlenecks with smaller buffers (down to half the \ac{BDP}) as it only reduces the sending rate by 0.3 but in return it causes a larger standing queue otherwise.
For Highspeed TCP as in Figure~\ref{fig:highspeed} the increase rate is the higher the larger the link bandwidth is (as it can be see by the slopes of the black and red lines). 
This improves the scalability but the feedback rate still depends on the average \ac{BDP}.
Further, the fixed decrease factor either underutilizes the link or causes a standing queue.
Scalable TCP as an MIMD scheme scales well, but often causes a very large standing queue, as it can be seen in Figure~\ref{fig:scalable}.
Figure~\ref{fig:htcp} displays the behavior of H-TCP.
It can clearly be seen that all high speed schemes have in this example scenario a higher oscillation rate than TCP NewReno as this it needed for scalability.
While Scalable TCP and H-TCP have a fixed and quite high feedback rate, the feedback rate of TCP SIAD can be configured.
In the shown scenario in Figure~\ref{fig:siad} $Num_{RTT}$ is set to 20 and therefore TCP SIAD induces a similar feedback rate as induced by H-TCP.
Further similar to H-TCP, TCP SIAD adapts its decrease factor to the queue size.
As H-TCP only implements a decrease factor between 0.3 and 0.5, it still causes a standing queue when the buffer is larger than the base \ac{BDP}.
In contrast for  TCP SIAD we removed this limitation. 
In addition TCP SIAD performs additional decreases from time to time due to our requirement to drain the queue completely.
This can be clearly seen for the red curve in Figure~\ref{fig:siad} where stronger dips occur.
Although we can observe a couple of strong decreases, Scalable Increase ramps up even quicker afterwards and thereby can avoid too much underutilization of the link.

%

\begin{figure}[t]
\centering
  \subfigure[Average queue fill level.]{
    \label{fig:qfill}
    \includegraphics[trim = 10mm 15mm 10mm 55mm, clip, width=0.45\textwidth]{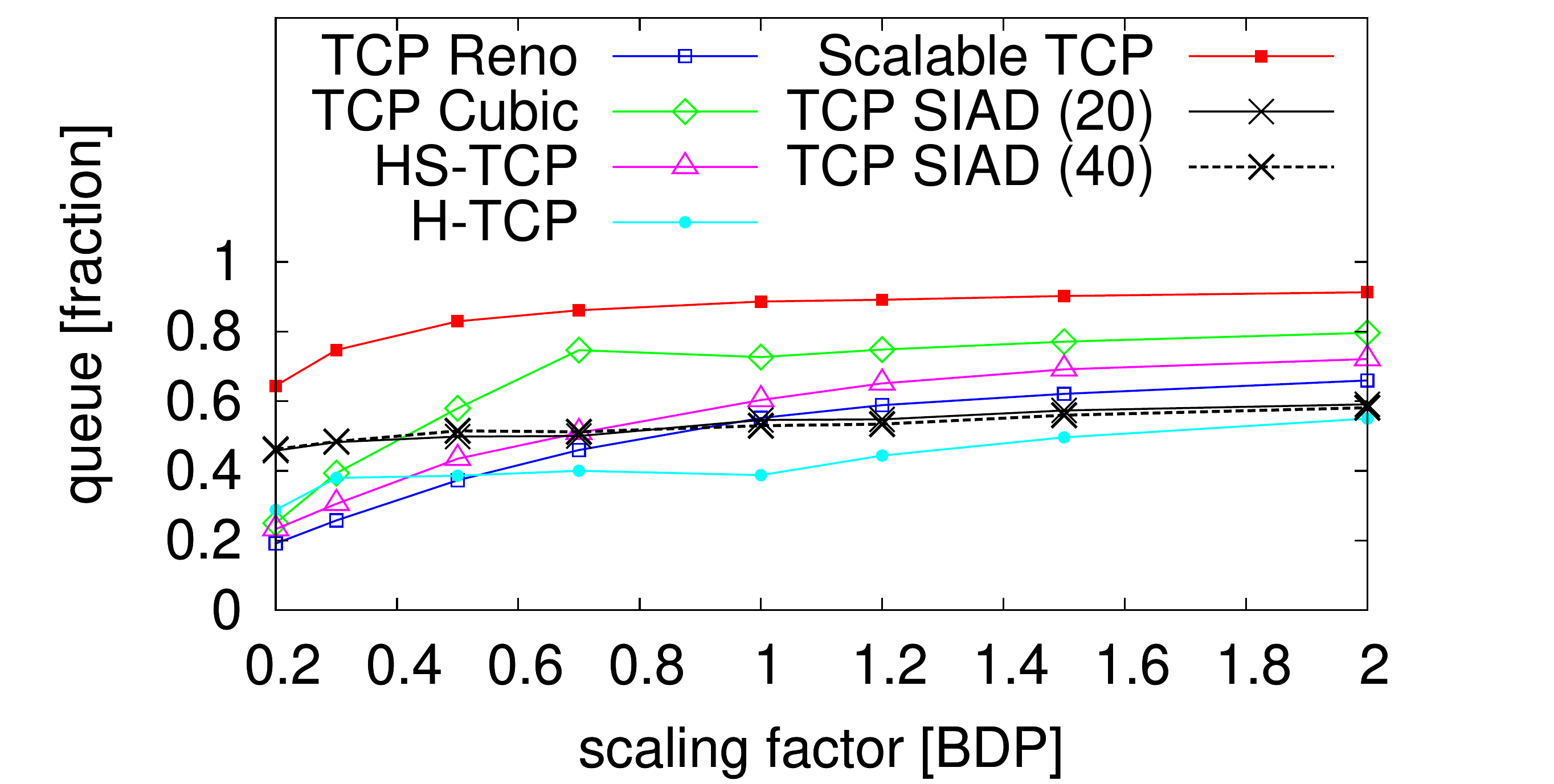}
    }
  \subfigure[Average link utilization.]{
    \label{fig:util}
    \includegraphics[trim = 10mm 15mm 10mm 55mm, clip, width=0.45\textwidth]{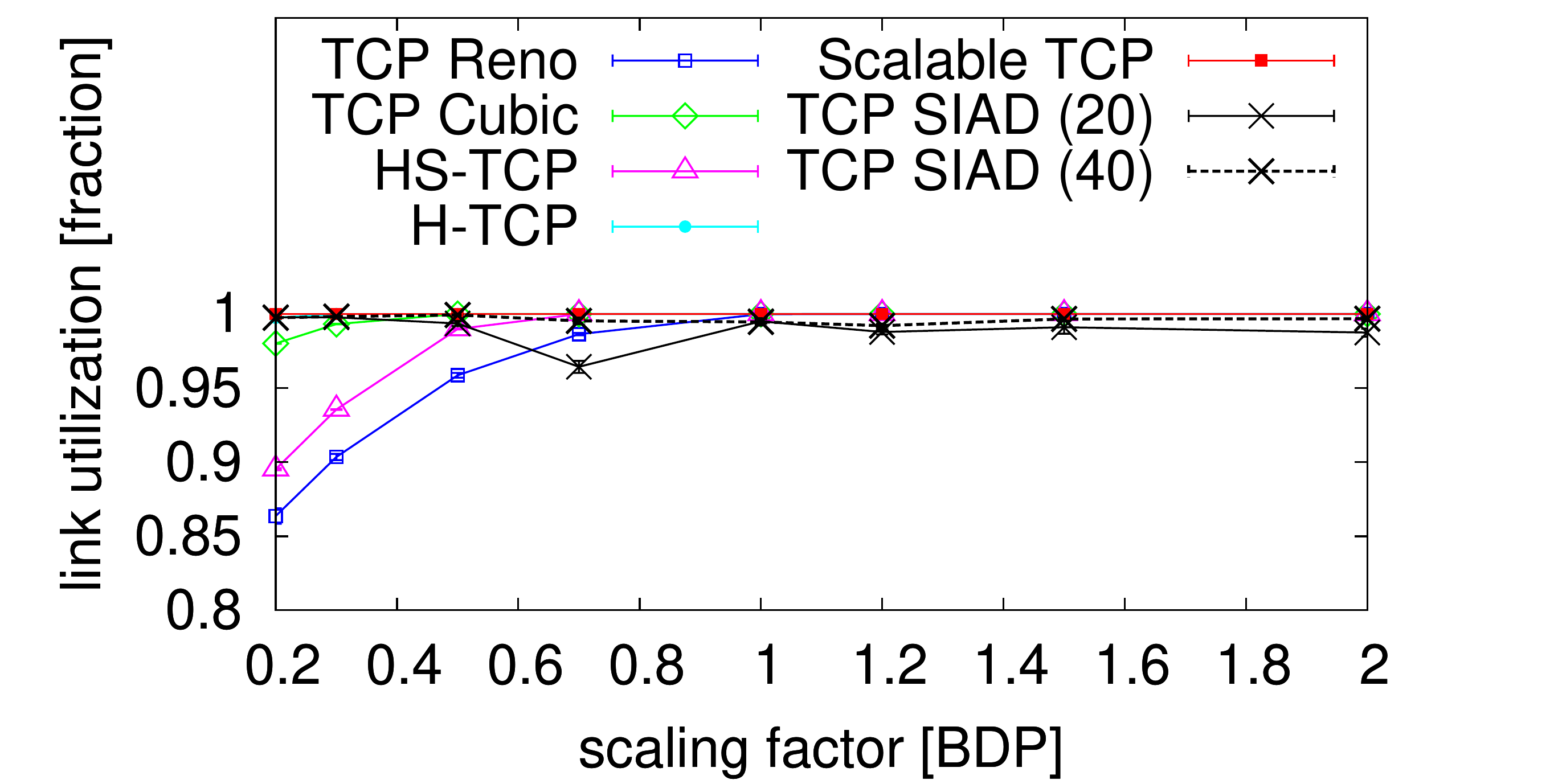}
    }
\caption{Single flow at 10\,Mbit/s.}
\end{figure}

\begin{figure}[t]
\centering
  \subfigure[Average loss rate.]{
    \label{fig:loss}
    \includegraphics[trim = 10mm 15mm 10mm 55mm, clip, width=0.45\textwidth]{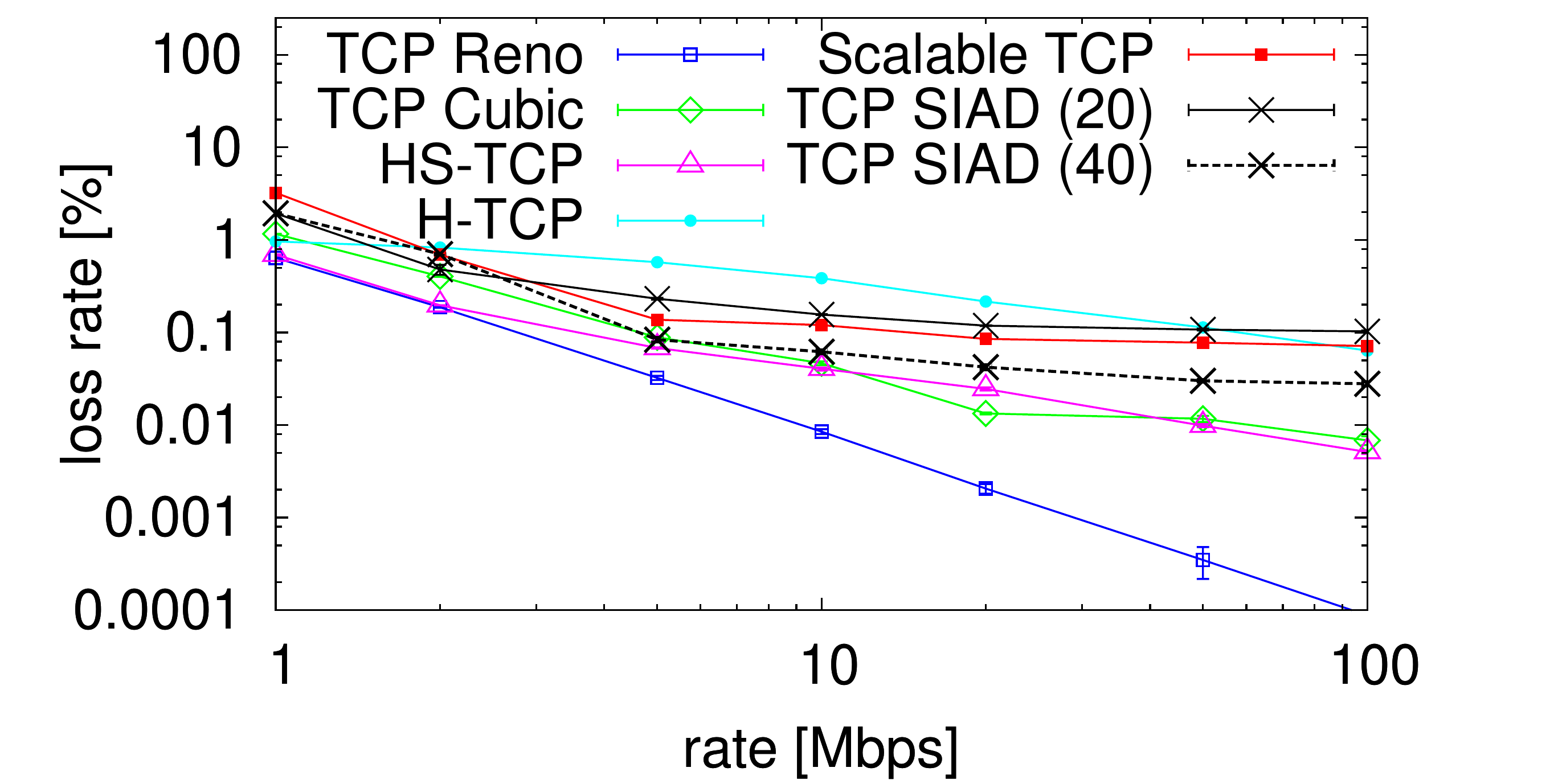}
    }
  \subfigure[Average loss event distance.]{
    \label{fig:lossFreq}
    \includegraphics[trim = 10mm 15mm 10mm 55mm, clip, width=0.45\textwidth]{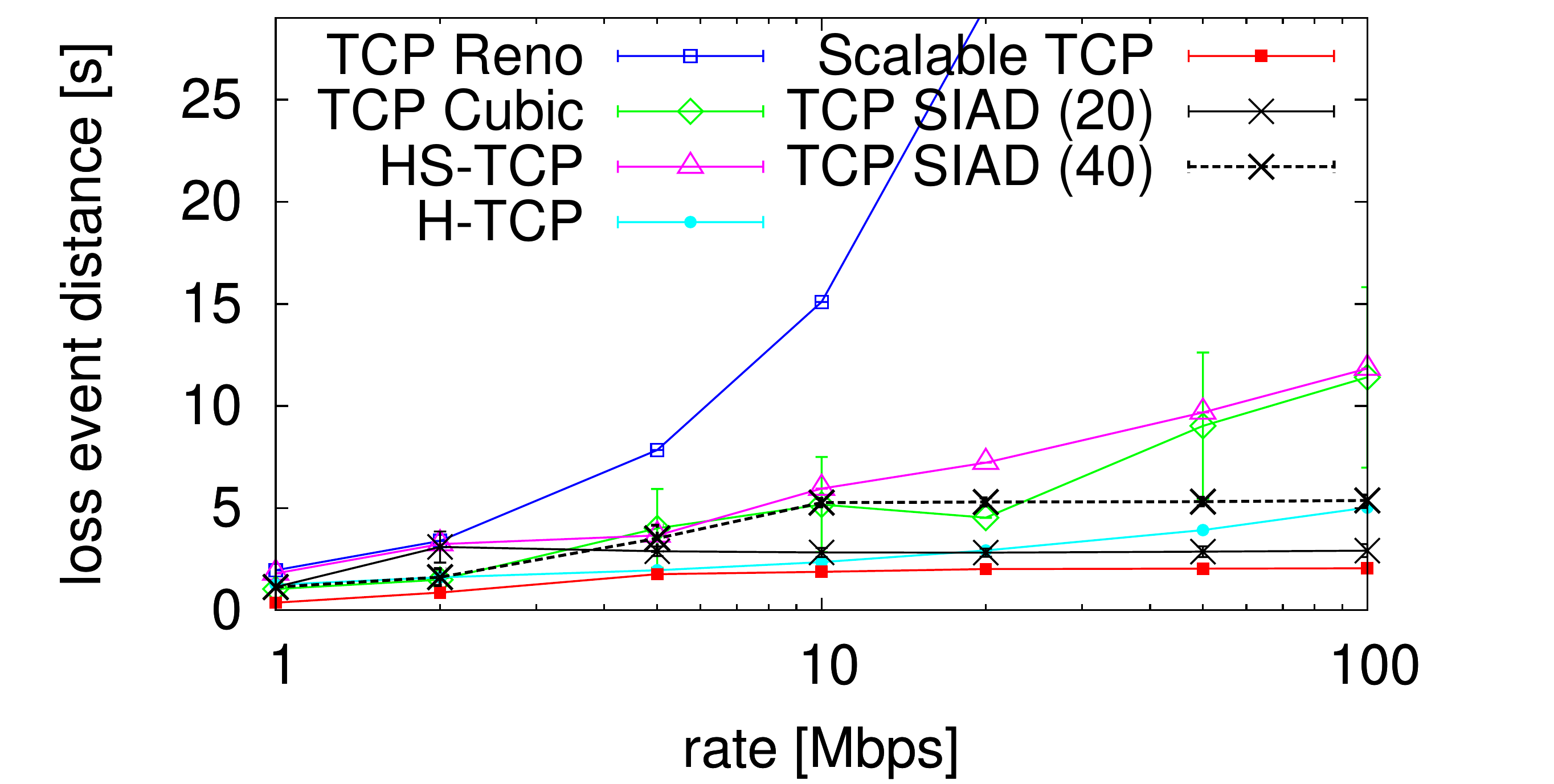}
    }
\caption{Single flow with queue size of 0.5*\ac{BDP}}
\end{figure}

In Figure~\ref{fig:qfill} and~\ref{fig:util} the average link utilization as well as the average queue fill are shown for simulation runs with a bandwidth of 10\,Mbit/s and different buffer sizes.
As expected, only Scalable TCP, H-TCP and TCP SIAD can utilize the link fully with very small buffer configurations. 
While TCP SIAD maintains an average queue fill level of about 50\% in all scenarios, for all other schemes the queue fill increases with the configured buffer size.
This leads to a standing queue and unnecessary additional delays. 
H-TCP reaches the same average queue fill as TCP SIAD in case of a buffer size of 1.5*\ac{BDP}.
In this scenario H-TCP already induces a standing queue (minimum queue fill is larger than 0), but as it increases its sending rate more than linearly, it simply stays longer at a lower queue fill level which reduces the average queuing delay.
The drawback of this increase function is the large overshoot and therefore a higher loss rate, as can be seen in Figure~\ref{fig:loss}.
We also considered different (not linear) increase functions for the design of TCP SIAD, but with the expectation of smaller queues in future networks the gain in latency does not justify the additional complexity.

In summary, TCP SIAD achieves high link utilization for all scenarios.
All other schemes only provide full utilization if the buffer is large enough (due to a standing queue).
In contrast, TCP SIAD is designed to rather slightly underutilize the link but drain the queue with every decrease. 
Therefore e.g. in the scenario with a maximum queue size of 0.7*BDP, TCP SIAD only achieves a slightly lower utilization than the other schemes (but still 96.4\%). 
In this scenario TCP SIAD frequently performs Additional Decreases as the minimum delay is not measured correctly.


While one goal of TCP SIAD is high utilization even with small queues, another goal is to maintain a given feedback rate and therefore scale with any network bandwidth.
In Figures~\ref{fig:loss} and~\ref{fig:lossFreq} we show the average loss rate and the average time between two loss events.
In this set-up the maximum queue size was configure to 0.5*\ac{BDP} while the link bandwidth is varied.
Even though all high speed proposals scale better than TCP Reno, only Scalable TCP and TCP SIAD maintain a feedback rate independent of the available bandwidth.
For Scalable TCP the feedback rate is fixed which induces for most Internet scenarios a high loss rate and large queuing delays.
In contrast TCP SIAD can be configured to maintain a given congestion event distance which also influences the loss rate. 
Note that for $Num_{RTT}=40$ and small \acp{BDP} the target feedback rate is exceeded, as the maximum congestion window is less than 40 packets and TCP SIAD has a minimum increase rate of 1 packet per \ac{RTT}.

%

\begin{figure}[b]
\centering
\includegraphics[trim = 20mm 15mm 30mm 17mm, width=0.35\textwidth]{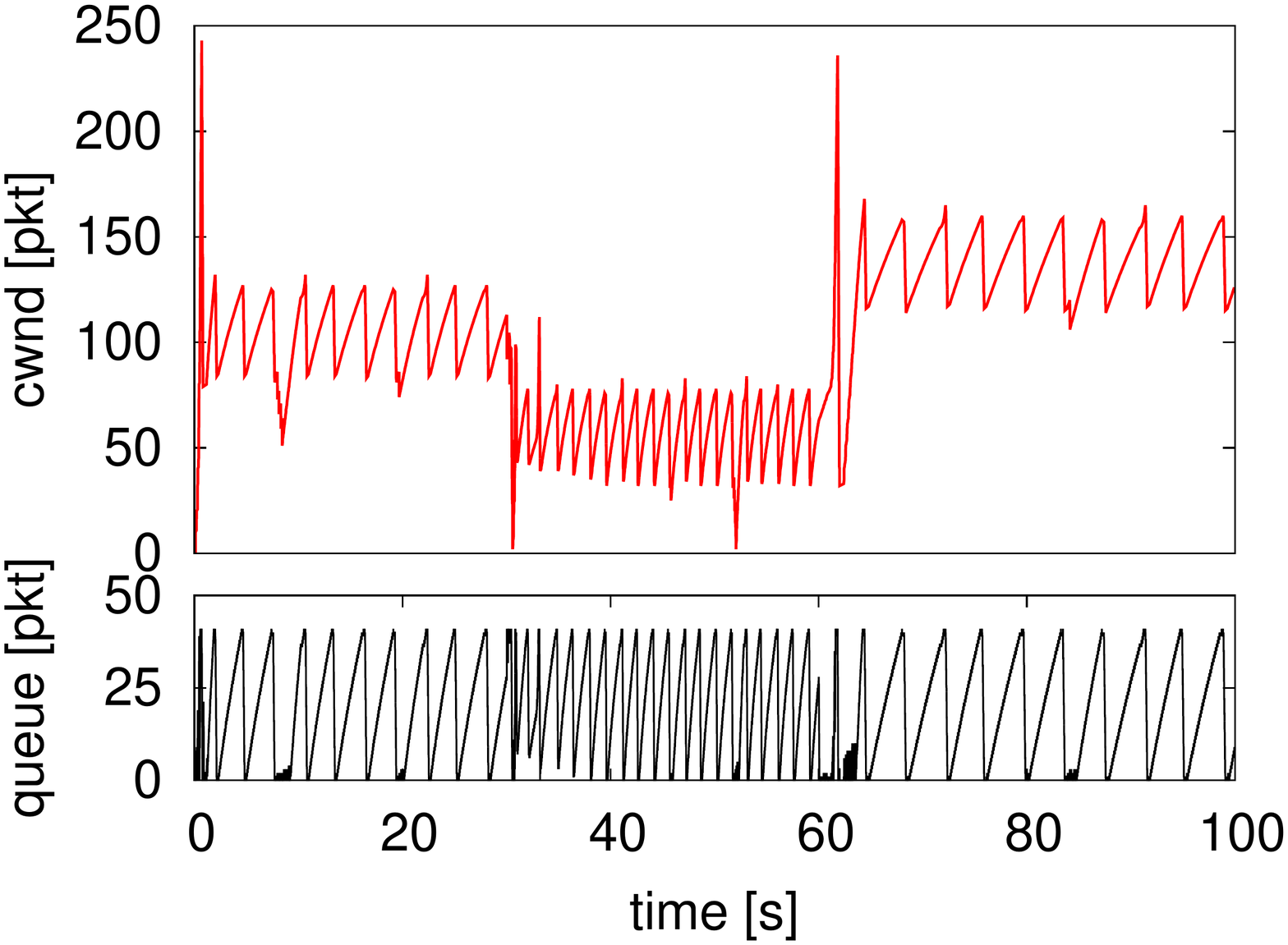}
\caption{Single TCP SIAD flow with changing base RTT.}
\label{fig:changingRTTs} 
\end{figure}

\vspace{0.2cm}
Further we demonstrate TCP SIAD's capability to adapt to changing \acp{RTT}.
Figure~\ref{fig:changingRTTs} shows one single TCP SIAD flow on a 10\,Mbit/s link with a bottleneck buffer size of 0.5 BDP. 
At beginning of the simulation the \ac{RTT} is 100\,ms, after 30\,s of simulation time the \ac{RTT} is changed to 40\,ms, and at 60\,s of simulation time the \ac{RTT} is increased to 140\,ms. 

It takes a few overshoots before TCP SIAD correctly adapts to a smaller \ac{RTT}. 
Even though the smaller base \ac{RTT} can be measured immediately after a decrease, also $incthresh$ and $trend$ have to adapt to a new maximum congestion window which is also smaller now. 
In contrast the adaptation to a larger value is completed as soon as one decrease with the old value is performed. 
Of course the old decrease value provokes a too large decrease but due to this too large decrease the link gets underutilized for a short time and the minimum delay can be updated immediately. 
Subsequently Fast Increase provides quick re-allocation of the link resources.

In addition we have evaluated the impact of \ac{RTT} variation on TCP SIAD.
E.g. with a random additional delay add-on between 0\,ms and 3\,ms on a 10\,Mbit/s link with 1 \ac{BDP} of buffering, TCP SIAD ($Num_{RTT}=20$), as all hybrid scheme, has a slightly lower utilization of 93.2\% (compared to 99.9\% for NewReno and 99.7\% for Tcp Cubic) but still a much higher utilization than H-TCP (82.8\%).
Further not that TCP SIAD's utilization increases for larger $Num_{RTT}$ values.

\begin{figure}[t]
\centering
\includegraphics[width=0.45\textwidth]{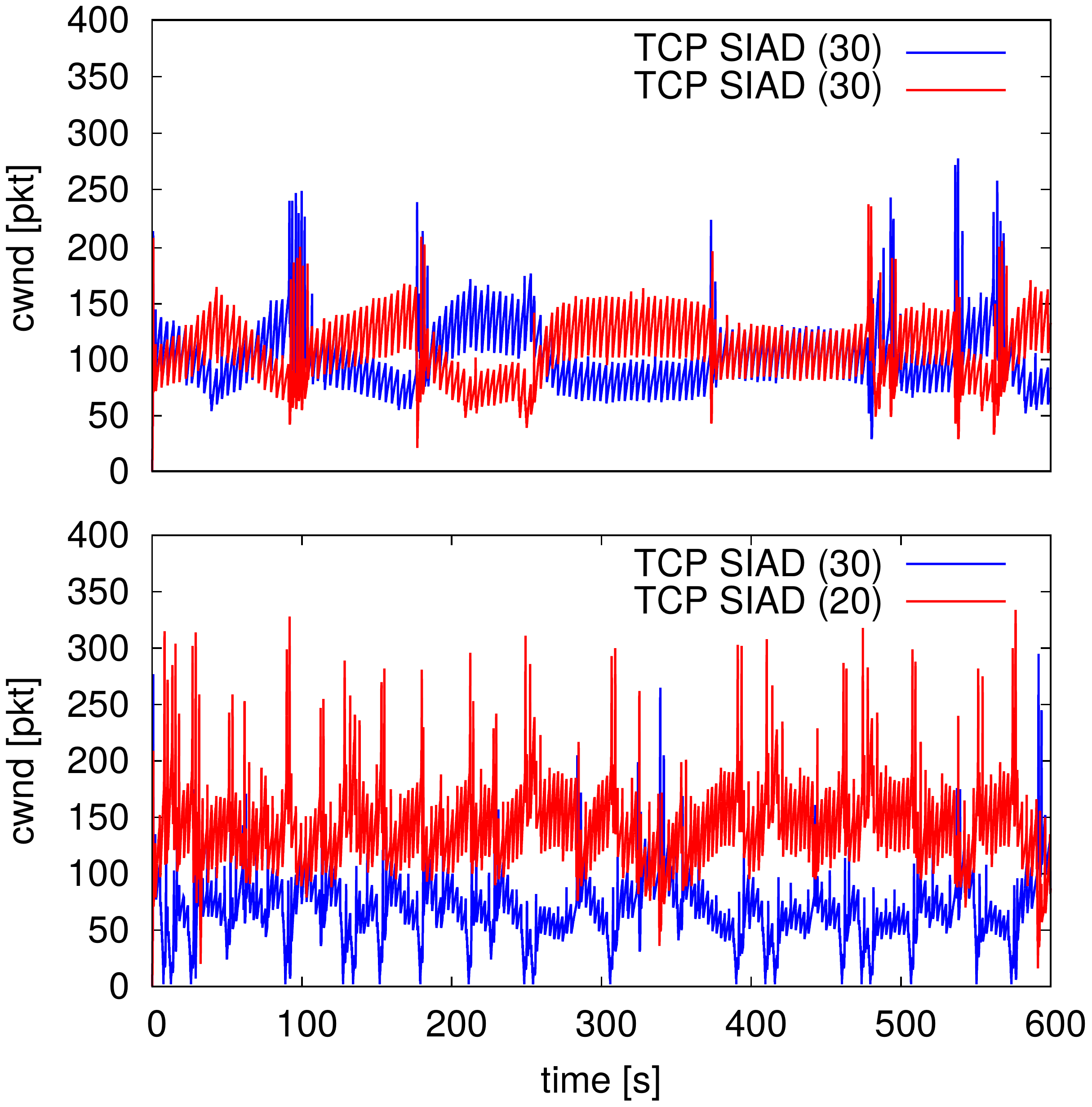}
\caption{Two TCP SIAD flows at 20\,Mbit/s.}
\label{fig:2siad}
\end{figure}

We conclude that TCP SIAD achieves the design goals on high link utilization, minimizing the average queuing delay, and a constant feedback rate to implement scalability in this initial steady state scenario with one flow.
These goals can be reached independent of the $Num_{RTT}$ configuration as shown for two example values of 20 and 40.
Further, we have performed the same set of simulations with two competing flows to assess the average link utilization and queue fill level for different buffer configurations as well as the average loss rate and loss event distance for different link speeds.
These simulations are not shown due to space limitation but provide similar results indicating that the respective design goals are reached.
In the following sections we focus on quick bandwidth allocation in dynamic scenarios and a configurable capacity sharing between multiple flows.

\subsection{Capacity Sharing and Convergence with Multiple Flows}\label{sec:eval:multipleFlows}

In this scenario we evaluate capacity sharing and convergence of multiple TCP SAID flows.
Figure~\ref{fig:2siad} shows two TCP SIAD flows on a 20\,Mbit/s link and a maximum queue size of 0.5*\ac{BDP}.
In the upper plot the two flows use the same $Num_{RTT}$ configuration value of 30.
The flows achieve an average rate of 9.61\,Mbit/s and 10.26\,Mbit/s, respectively, and a link utilization of 99.85\,\% and average queue fill faction of 0.51. 
In the lower plot one flow uses a different $Num_{RTT}$ of 20, instead. 
This flow has an average rate of 13.07\,Mbit/s while the other one achieves 6.86\,Mbit/s with a link utilization of 99.7\% and average queue fill of 0.44.
This demonstrates well that TCP SIAD's configuration parameter $Num_{RTT}$ can be used to influence the capacity sharing between competing flows.
A higher layer control loop can use this parameter to adjust its share during the transmission and thereby better cope with application requirements like a minimum sending rate.
However, we assume all flows start with a default (per-interface) value that is chosen such that it can provide about equal sharing for a small numbers of competing (traditional TCP) flows at the given access link speed. 
Only if the application requirements cannot be met and therefore a service would otherwise not be usable, this parameter should be adapted.

\begin{figure}[t]
\centering
\includegraphics[trim = 10mm 15mm 10mm 15mm, clip, width=0.45\textwidth]{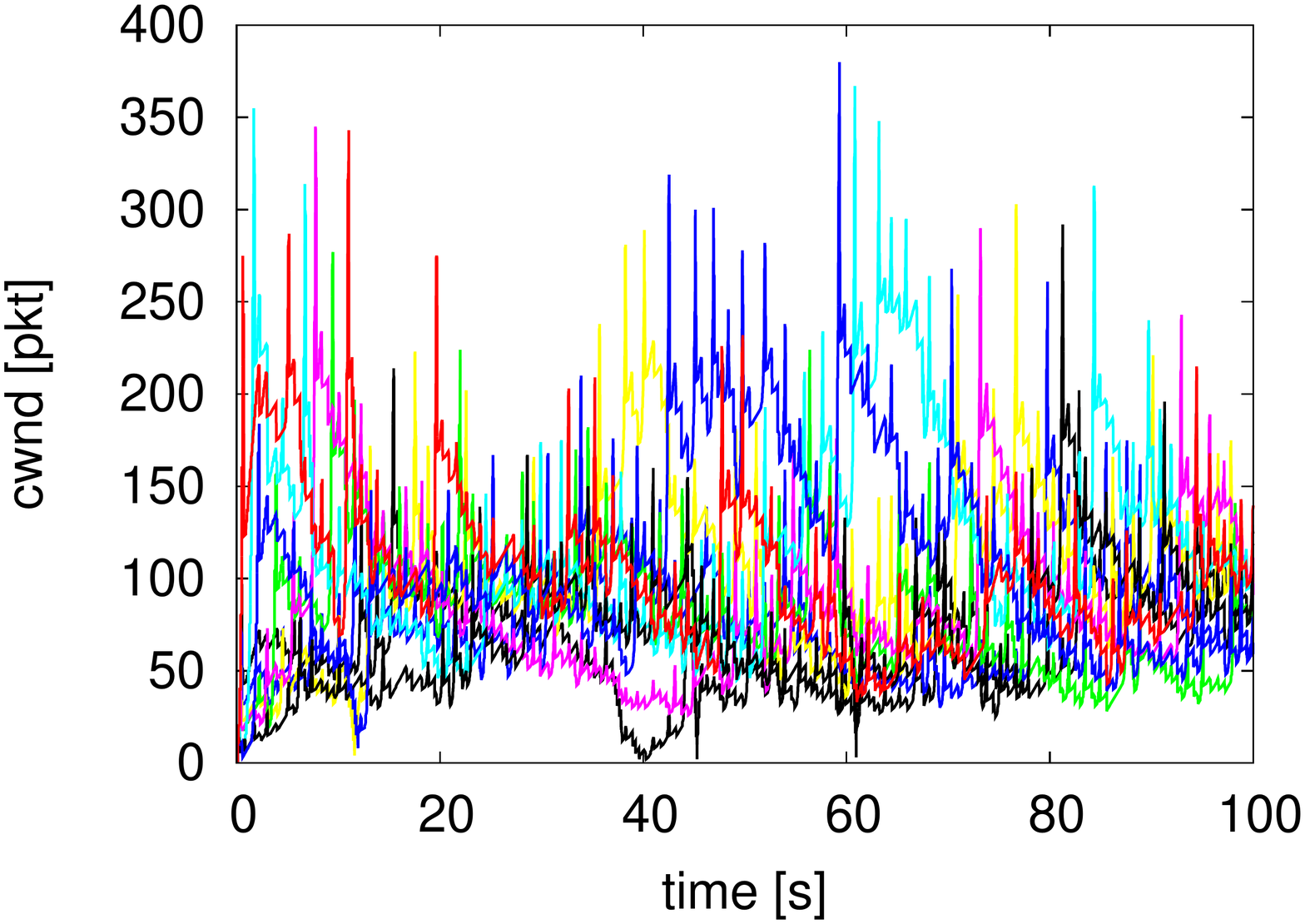}
\caption{10 TCP SIAD flows at 100\,Mbit/s.}
\label{fig:10siad-100}
\end{figure}

We also evaluated a two-flow scenario where the competing TCP SIAD flows have different base \acp{RTT} of 100\,ms and 200\,ms on a 20\,Mbit/s link and a buffer size of 1*\ac{BDP} based on an average \ac{RTT} of 150\,ms.
Using the same $Num_{RTT}$ value for both flows, TCP SIAD implements RTT-unfairness as most common schemes.
The two flows achieve an average rate of 15.63\,Mbit/s and 4.27\,Mbit/s.
Even though addressing RTT-unfairness is not a desired goal, TCP SIAD's aggressiveness can also be set based on a absolute time base. 
By using e.g. $Num_{MS}=4000$ both flows achieve a more equal sharing of 10.93\,Mbit/s and 8.95\,Mbit/s.
We proposed the use of $Num_{RTT}$ for the Internet as often a wide variety of \ac{RTT} can be found.
In this case it makes sense to adapt faster and potentially even get a larger share for connections that are close by, e.g. connections to a local CDN. 
In other scenarios, e.g. in a data centers, it might be more sensible to set the configuration parameter on an absolute time base to have all entities behaving similar.

In Figure~\ref{fig:10siad-100}, ten TCP SIAD flows are competing on a larger bandwidth link of 100\,Mbit/s link and a very small buffer size of 0.1*\ac{BDP}.
The red bold line shows the congestion window of the flow that puts the first packet in the queue.
While this flow at the beginning grabs slightly more capacity, it still has (only) an average rate of 10.67\,Mbit/s over the total simulation run of 580s.
Even though flows have quite diverging shares on a short time scale, the capacity is shared about equally on a longer time scale.
In this scenario the smallest share is 9.32\,Mbit/s compared to a maximum of 11.04\,Mbit/s.
This is similar to the behavior of other high speed as larger aggregates of flows most often are not synchronized.

\begin{figure}[t]
\centering
\includegraphics[trim = 20mm 15mm 30mm 17mm, clip, width=0.45\textwidth]{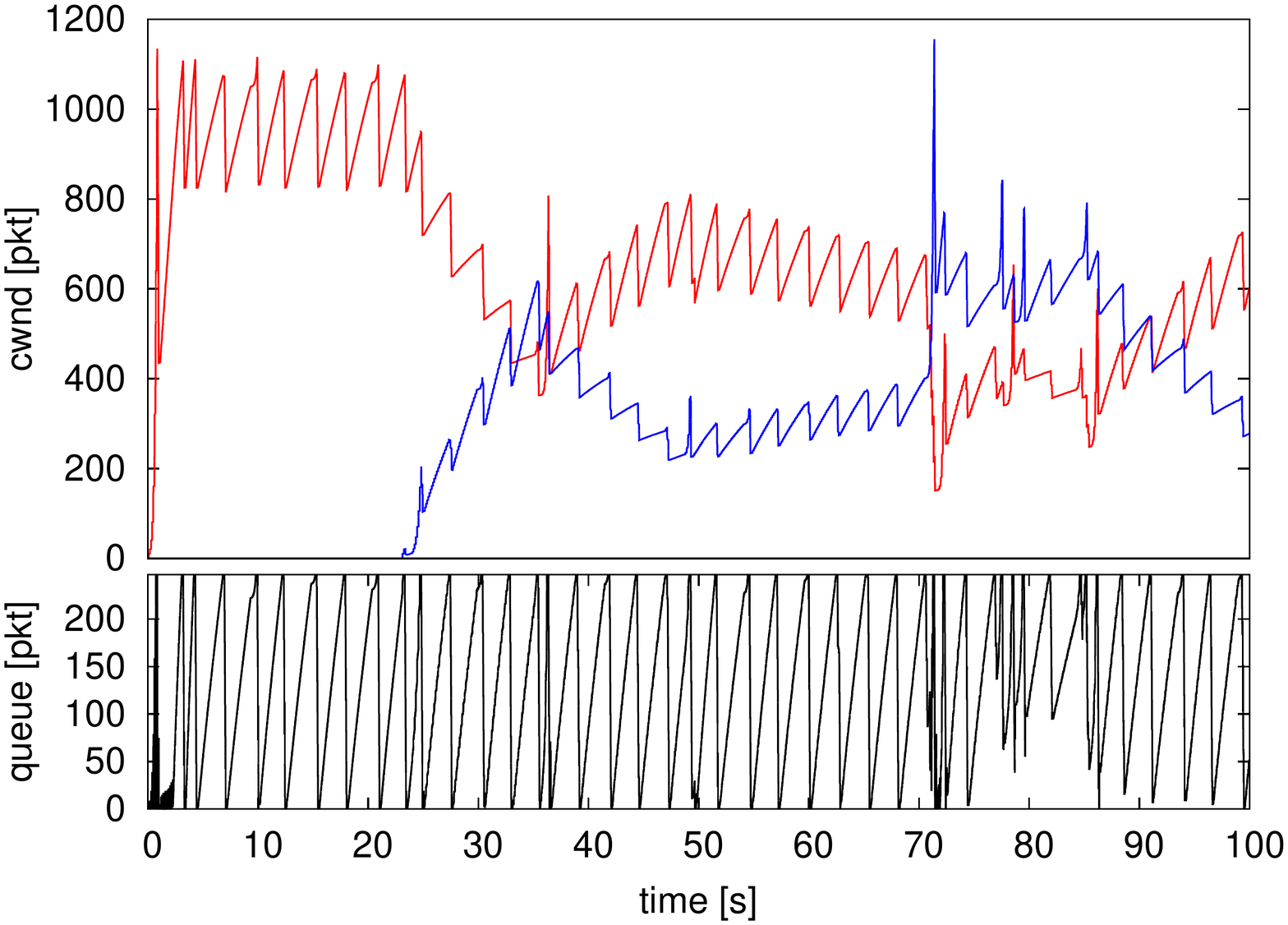}
\caption{Two TCP SIAD flows at 100\,Mbit/s.}
\label{fig:2siad-100}
\end{figure}

Figure~\ref{fig:2siad-100} shows a scenario with 100\,Mbit/s and a buffer size of 0.3*\ac{BDP} (which still covers more than 240 packets but induces only 30\,ms of maximum queuing delay).
A second flow starts after 60 seconds of simulation time and stops again at 180 seconds.
Note that the convergence time not only depends on the increase rate of the newly starting flow but also strongly on the decrease factor of the existing flow and therefore the buffer size:
Only if the first flow gets out of the way, the second flow can grab some capacity.
Even though the decrease factor is quite small in this scenario, both flows still converge after a reasonable amount of time.

\begin{figure}[t]
\centering
\includegraphics[trim = 10mm 15mm 10mm 52mm, clip, width=0.45\textwidth]{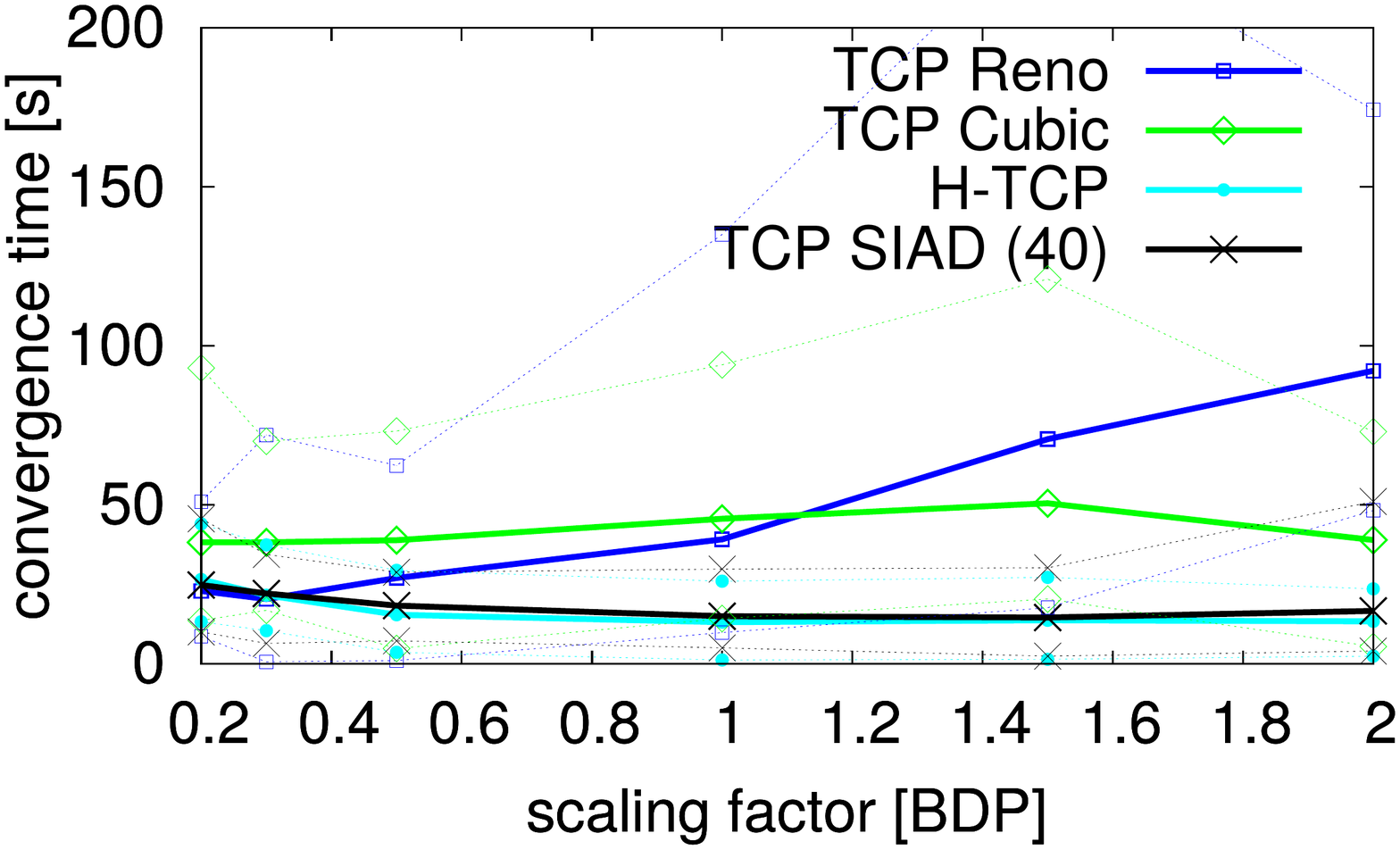}
\caption{Average, min, max convergence time.}
\label{fig:convergenceTime}
\end{figure}

Figure~\ref{fig:convergenceTime} shows the average, minimum, maximum convergence time over the buffer size on a 20\,Mbit/s link.
Each point represents 20 simulation runs.
In each run a second flow starts at a different time; at each full second between 20\,s and 39\,s.
The minimum and maximum is only shown to illustrate the large variation within our 20 sample measurements.
Due to the strongly different congestion epoch length of each scheme it is hard to get comparable and statistical relevant measurements. 
However, strong differences in the general behavior can be demonstrated by this evaluation:
The convergence time of TCP NewReno increases strongly with the buffer size.
Further, TCP Cubic is rather slow due to the concave increase behavior that keeps the sending rate for a long time at the estimated target rate.
In contrast, TCP SIAD enters Fast Increase after reaching $incthresh$ and reallocates the available capacity very quickly.
H-TCP achieves a similar convergence than TCP SIAD but induces more loss due to the aggressive increase behavior.

We conclude that TCP SIAD achieves a reasonable convergence time in comparison to other high speed schemes and outperforms TCP Cubic.
Further the aggressiveness of TCP SIAD can 
influence 
the capacity sharing between competing flows. 
In the following section we further demonstrate how this principle can be used with non-SIAD cross traffic.

%

%

\subsection{Competing with Non-SIAD Flows and Cross Traffic}

TCP SIAD is explicitly not designed for TCP-friendliness.
However, 
TCP SIAD can still be configured to achieve equal sharing which is shown in the following scenarios with TCP NewReno and TCP Cubic cross traffic.
In addition we afterwards demonstrate TCP SIAD's high robustness to loss e.g. induced by short flow cross traffic.

\begin{figure}[t]
\centering
\includegraphics[width=0.45\textwidth]{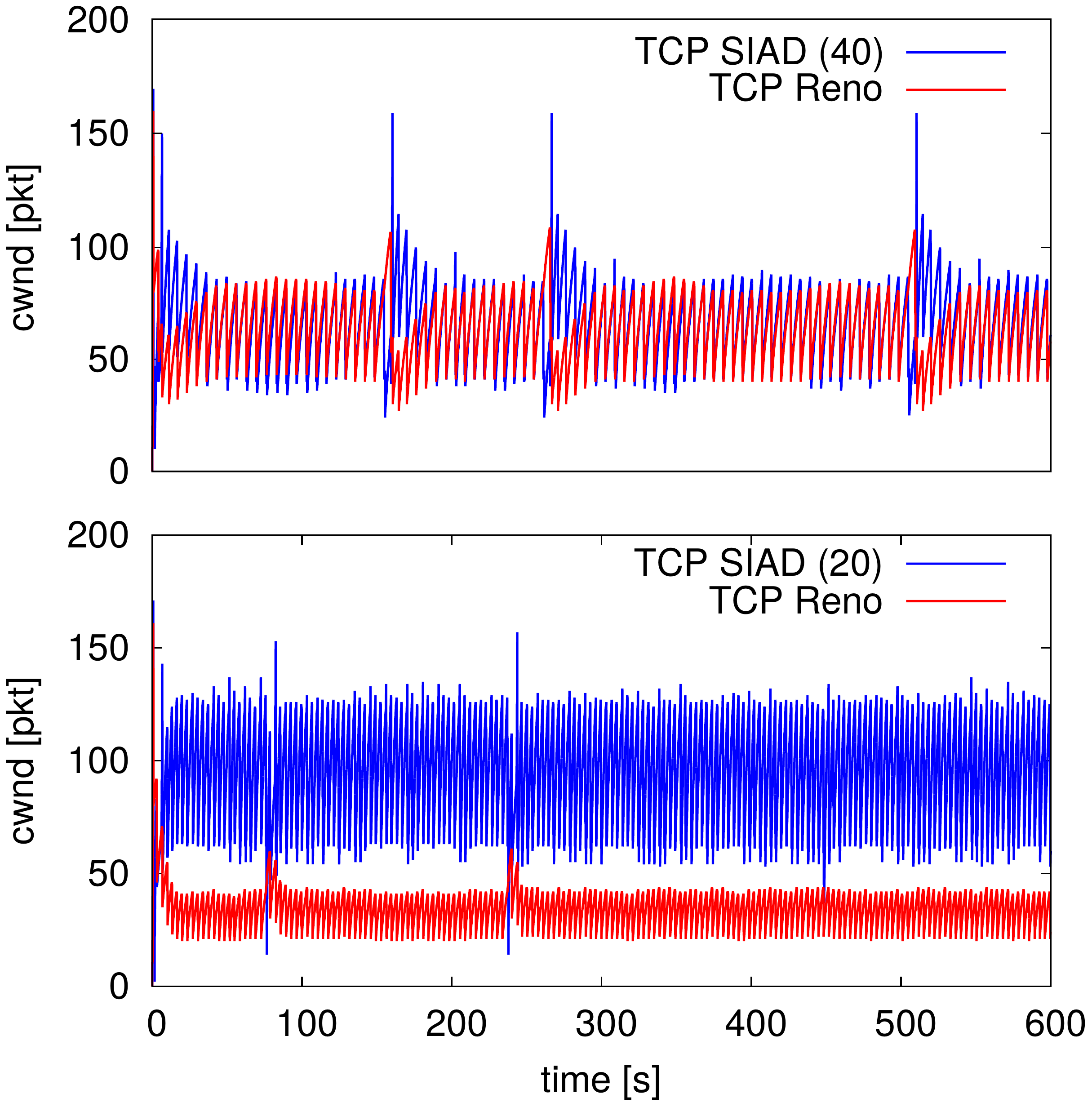}
\caption{TCP NewReno cross traffic at 10\,Mbit/s.}
\label{fig:crossReno}
\end{figure}

In Figure~\ref{fig:crossReno} one TCP SIAD and one TCP NewReno flow share the same bottleneck with a link bandwidth of 10\,Mbit/s and a maximum queue size of 1*\ac{BDP}.
For this simulation we disabled delayed acknowledgements for the NewReno flow to reach an increase rate of 1 packet per \ac{RTT}, as the current Linux implementation does not compensate the lower ACK rate. 
Otherwise we could never reach equal sharing with TCP NewReno, as TCP SIAD implements a minimum increase of one packet per \ac{RTT}.
In the upper plot TCP SIAD is configured with $Num_{RTT}$ set to 40 which is about half the \ac{BDP} and therefore the oscillation size of TCP NewReno when it has half the bandwidth allocated.
It can be seen that both flows share the link about equally; more precisely the NewReno flow get 5.03\,Mbit/s and TCP SIAD 4.97\,Mbit/s.
In the lower plot $Num_{RTT}$ is set to 20, and therefore TCP SIAD gets a larger share of 7.37\,Mbit/s, while the TCP NewReno flow has an average rate of 2.62\,Mbit/s.

\begin{figure}[t]
\centering
\includegraphics[width=0.45\textwidth]{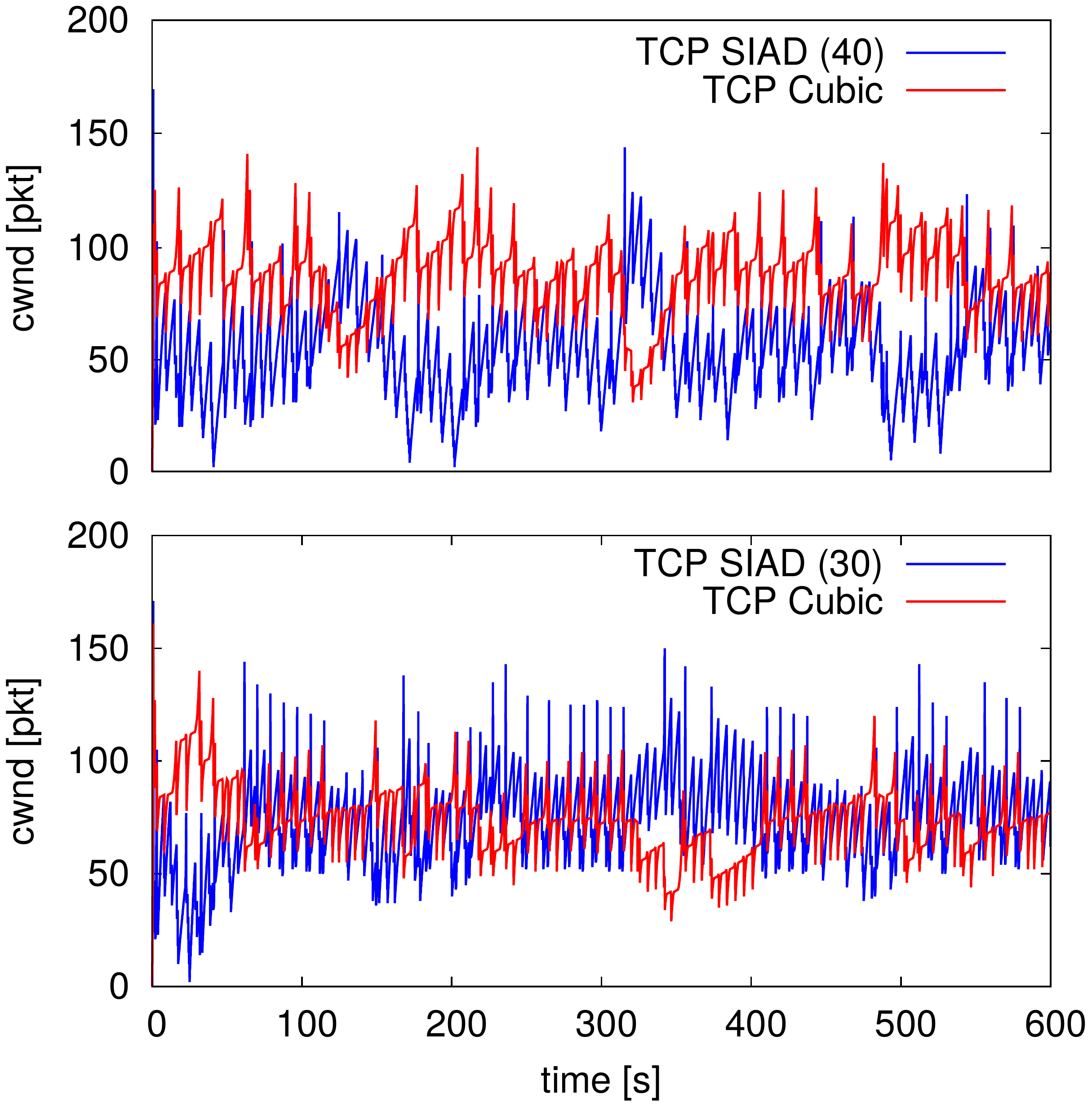}
\caption{TCP Cubic cross traffic at 10\,Mbit/s.}
\label{fig:crossCubic}
\end{figure}

In Figure~\ref{fig:crossCubic} 
TCP Cubic with delayed ACKs is used instead.
We can see that TCP Cubic is more aggressive and therefore gets the larger share. 
While TCP SIAD with $Num_{RTT}=40$ has an average rate of 3.77\,Mbit/s only, the TCP Cubic flow sends with 6.23\,Mbit/s on average.
If we set the $Num_{RTT}$ to 30, as shown in the lower plot, we can again achieve about equal sharing with 4.98\,Mbit/s for the TCP SIAD flow compared to 5.02\,Mbit/s for the TCP Cubic flow.
In a more controlled environment than the Internet this mechanism can be used to maintain a given capacity sharing as well as equal sharing.
However, in the distributed congestion control system in the Internet every single endpoint can simply use a more aggressive scheme such as TCP Cubic or TCP SIAD with a small $Num_RTT$ value.
Therefore fairness must be induced by the network instead, as explained in the motivation.

\begin{figure}[t]
\centering
\includegraphics[trim = 10mm 15mm 10mm 55mm, clip, width=0.45\textwidth]{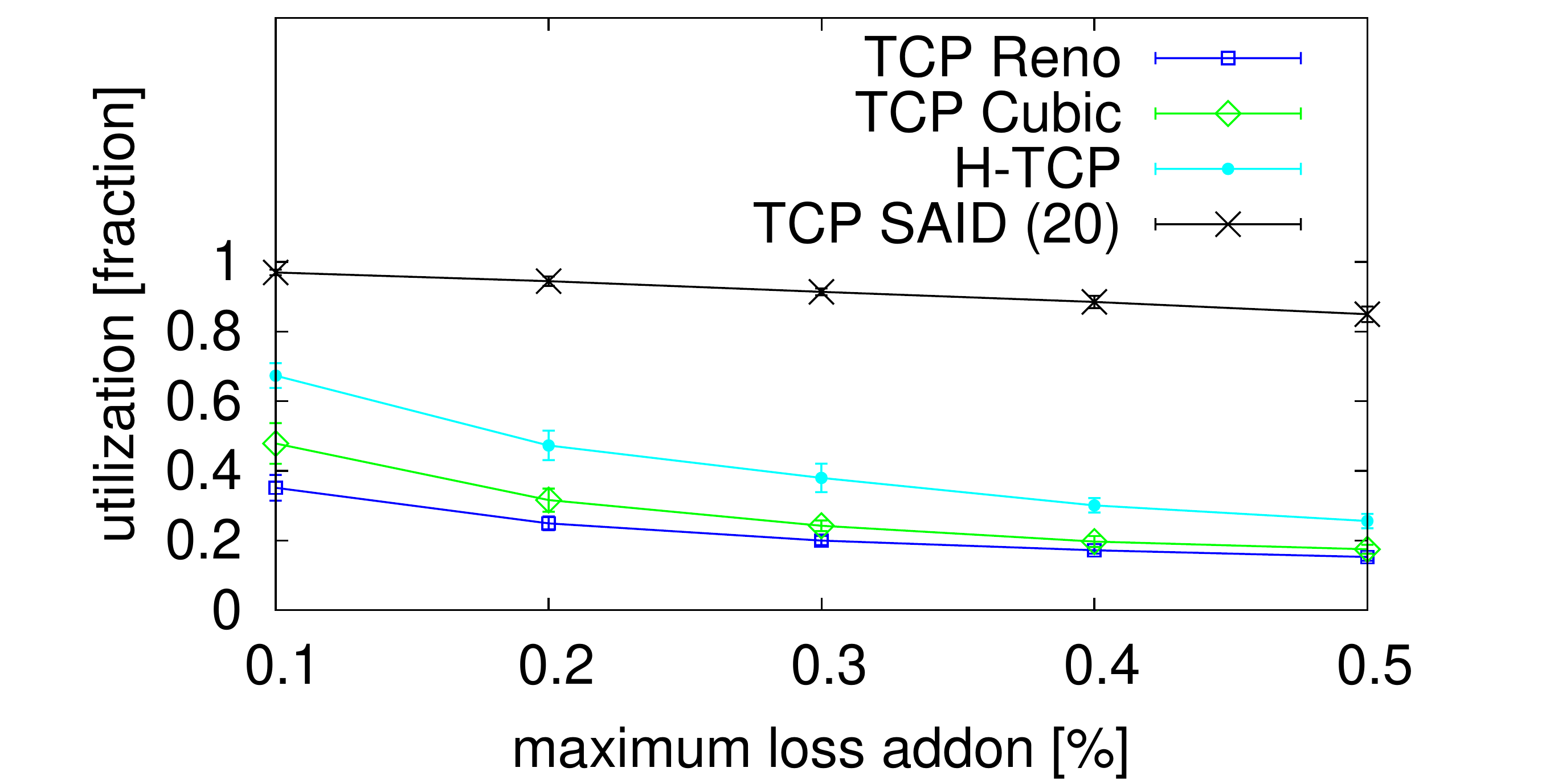}
\caption{Average link utilization at 10\,Mbit/s with 0.5*BDP buffering and random loss.}
\label{fig:randomLoss}
\end{figure}

Moreover, we performed further simulations in various traffic scenarios such as with additional random loss, short flow cross traffic, or multiple bottlenecks (no shown) to evaluate TCP SIAD in a wide range of extreme scenarios as they potentially can occur in the Internet.
Figure~\ref{fig:randomLoss} show the link utilization of one flow using different congestion control with additional, non-congestion related random loss between 0.2\% and 0.5\% on a 10\,Mbit/s link with 0.5*\ac{BDP} of buffering.
While TCP SIAD (e.g. with $Num_{RTT}=20$) can reach a utilization of 84.96\,\% -- 94.47\,\% due to Adaptive Decrease, all other schemes including H-TCP cannot utilize the link very well.
TCP NewReno only utilizes the link 15.25\,\% -- 24.93\,\%, TCP Cubic 17.46\,\% -- 31.6\,\%, and H-TCP 25.58\,\% -- 47.29\,\%.
In the more realistic case with competing short flow cross traffic bursts e.g. with 300\,KBytes of data and an uniformly distributed Inter-Arrival Time (IAT) between 2 and 3 seconds on a 10\,Mbit/s link with 0.5*\ac{BDP} of buffering, 
TCP SIAD ($Num_{RTT}=20$) achieves a link utilization of 93.64\,\% while TCP NewReno only reaches 59.37\,\% and TCP Cubic 89.69\,\%.

TCP SIAD's stable capacity sharing and its high robustness to loss shows the ability to cope with a wide range of Internet scenarios which is essential for congestion control design.

\section{Conclusion and Outlook}
In this paper, we presented a new TCP congestion control scheme, called TCP SIAD, supporting high speed and low latency 
by implementing \textit{Scalable Increase Adaptive Decrease}.
Scalable Increase dynamically determines the linear per \ac{RTT} increase step $\alpha$ such that a target congestion window value $incthresh$ is reached in a configurable number of \acp{RTT} to provide full scalability. 
In addition, Scalable Increase introduces a configuration parameter $Num_{RTT}$ that can be used by a higher layer control loop to impact TCP SIAD's aggressiveness. 
Adaptive Decrease calculates the decrease factor $\beta$ on congestion notification such that the queue just empties 
to always keep link utilization high, independent of the configured network buffer size.
This approach is similar to H-TCP, but 
due to its configurable aggressiveness,  TCP SIAD most often induces less lost by overshooting less.
Moreover, the combination of Scalable Increase and Adaptive Decrease 
allows us to not limit the decrease factor to a certain range, other than H-TCP does, and therefore to cope with small buffers that reduce the maximum delay as well as to minimize the average delay by avoiding a standing queue with large buffers.
Finally, we introduce a Fast Increase phase that quickly grabs newly available capacity by implementing a similar increase behavior as in Slow Start when the congestion window grows above the Linear Increment threshold $incthresh$.
We did not change the start-up behavior itself, namely Slow Start, as this is an research area on its own.
Both, Slow Start and Fast Increase, could potentially be enhanced when further extending TCP SIAD.
However, we argue that the introduction of a Fast Increase phase is essential for any high speed congestion control algorithm.

Simulative evaluation based on our Linux implementation has shown that TCP SIAD always induces a fixed feedback rate as well as achieves full utilization and avoids a standing queue on links with different bandwidth independent of the network buffer size.
This demonstrates the viability of the basic SIAD principle to provide scalability as well as low latency support.
We have further demonstrated TCP SIAD's  
high robustness to loss e.g when competing with short flow traffic bursts.
We have shown that TCP converges reasonable fast and faster than TCP Cubic. 
Moreover, TCP SIAD is able to achieve a stable capacity share between competing flows using the same or different congestion control.
Therefore, the configuration parameter $Num_{RTT}$ either directly defines the feedback rate (when alone on the bottleneck link) or the share of the capacity between the competing flows.

Even though equal sharing can be achieved with TCP SIAD, we do not see this as a requirement in today 's Internet anymore as long as every flows is able to grab sufficient capacity to fulfill the application's requirements.
However, we expect in most cases, and always at the beginning of a connection, that a default value for $Num_{RTT}$  is used.
This default value should be derived from typical Internet usage scenarios, e.g. with respect to the feedback rate that NewReno-like congestion control 
induces in a comparable scenario.
Additionally, some applications might implement a higher layer control loop to dynamically adapt the aggressiveness, 
e.g. for real-time video, where the service needs a certain minimum rate to work at all.
In this case the aggressiveness can be increased to grab a larger share 
at the cost of higher congestion,  at least for a certain time (until the congestion allowance at the policer is consumed). 
A detailed study of such an higher layer control loop is not part of this work as it strongly depends on the specific application.
However, we deem that it is important to have this configuration possibility as a basis for deployment of per-user congestion policing.
Note, there currently is no mechanism that stops single flows or users to use a more aggressive congestion control and thereby push away other traffic, as already happening when using TCP Cubic, the default configuration in Linux.
That means, fairness 
cannot be addressed solely by congestion control in a highly distributed system such as the Internet.
In summary, to better support emerging application that require low latency, both need to change, the buffer management in the network to reduce the maximum queuing delay as well as the congestion control in the end host to still achieve high link utilization.
A congestion control scheme like TCP SIAD 
allows network operators to configure smaller buffers and thereby provide low latency without causing network underutilization.
If the network operator does the first step and simply reduces its buffer sizes, this could initially lead to a lower link utilization but consequently incentivizes end-users to switch to a more appropriate congestion control like TCP SIAD that can fully utilize the link.
However, even with today's large buffer the use of TCP SIAD provides full scalability, fast bandwidth allocation, high robustness to loss, and a controlled average queuing delay by avoiding standing queues and therefore gives a deployment incentive to end-users to do the first step.

\balance
\bibliographystyle{abbrv}
\bibliography{../../bibtex/congestion_control}
\begin{acronym}[]
\acro{ABC}{Appropriate Byte Counting}
\acro{ACK}{acknowledgment}
\acro{AIMD}{Additive Increase Multiplicative Decrease}
\acro{API}{Application Programming Interface}
\acro{AQM}{Active Queue Management}
\acro{BDP}{Bandwidth-Delay-Product}
\acro{CBR}{Constant Bit Rate}
\acro{cwnd}{congestion window}
\acro{CoDel}{Controlled Delay}
\acro{DCTCP}{Data Center TCP}
\acro{ECN}{Explicit Congestion Notification}
\acro{EWMA}{Exponetially Weighted Moving Average}
\acro{FIFO}{First In First Out}
\acro{IAT}{Inter Arrival Time}
\acro{IETF}{Internet Engineering Task Force}
\acro{IP}{Internet Protocol}
\acro{IW}{Initial Window}
\acro{MIMD}{Multiplicative Increase Multiplicative Decrease}
\acro{MPLS}{Multi-Protocol Label Switching}
\acro{MTU}{Maximum Transmission Unit}
\acro{OS}{Operating System}
\acro{OWD}{One-Way-Delay}
\acro{PIE}{Proportional Integral Controller Enhanced}
\acro{PRR}{Proportional Rate Reduction}
\acro{RED}{Random Early Detection}
\acro{rmcat}{RTP Media Congestion Avoidance Techniques}
\acro{RTO}{Retransmission Time-Out}
\acro{RTT}{Round Trip Time}
\acro{RTTM}{Round-Trip Time Measurement}
\acro{SACK}{Selective Acknowledgment}
\acro{SEQ}{sequence number}
\acro{SCTP}{Stream Control Transmission Protocol}
\acro{TCP}{Transmission Control Protocol}
\acro{TFRC}{TCP-friendly Rate Control}
\acro{TSOpt}{TCP Timestamp Option}
\acro{UDP}{User Datagram Protocol}
\acro{QoE}{Quality of Experience}
\acro{QoS}{Quality of Service}
\acro{VM}{Virtual Machine}
\end{acronym}

\label{lastpage}
\end{document}